\documentclass[fleqn,usenatbib]{mnras}

\usepackage[utf8]{inputenc}
\usepackage[T1]{fontenc}
\usepackage{ae,aecompl}

\usepackage{graphicx}	% Including figure files
\usepackage{amsmath}	% Advanced maths commands
\usepackage{amssymb}	% Extra maths symbols
\usepackage{orcidlink}
\usepackage{newtxtext,newtxmath}

\newcommand{\tdep}{$\mathsf{\uptau_{depl}}$}

\title[Supermassive Starforming Clumps in LARS 8]{Unveiling the gravitationally unstable disc of a massive star-forming galaxy using NOEMA and MUSE\thanks{
This work is based on observations carried out under project numbers W16BS and E16AG with the IRAM NOEMA interferometer. IRAM is supported by INSU/CNRS (France), MPG (Germany), and IGN (Spain). Based on observations collected at the European Southern Observatory under ESO programme 0101.B-0703(A).}}

\author[Puschnig, Hayes, Agertz, Emsellem, Cannon...]{Johannes Puschnig,$^{1,2}$\thanks{johannes@jpuschnig.com \orcidlink{(0000-0003-1111-3951)}}
Matthew Hayes,$^{2}$
Oscar Agertz,$^{3}$
Eric Emsellem,$^{4}$
John M. Cannon,$^{5}$\newauthor
Jens Melinder,$^{2}$
G\"oran \"Ostlin,$^{2}$
Christian Herenz,$^{6}$
Veronica Menacho$^{2}$
\\
$^{1}$Universit\"at Bonn, Argelander-Institut f\"ur Astronomie, Auf dem H\"ugel 71, D-53121 Bonn, Germany\\
$^{2}$Department of Astronomy, Oskar Klein Centre, Stockholm University, AlbaNova University Centre, SE-106 91 Stockholm, Sweden\\
$^{3}$Lund Observatory, Department of Astronomy and Theoretical Physics, Lund University, Box 43, SE-221 00 Lund, Sweden\\
$^{4}$European Southern Observatory, Karl-Schwarzschild-Str. 2, D-85748 Garching, Germany\\
$^{5}$Department of Physics \& Astronomy, Macalester College, 1600 Grand Avenue, Saint Paul, MN 55105, US\\
$^{6}$European Southern Observatory, Av. Alonso de C\'ordova 3107, 763 0355 Vitacura, Santiago, Chile
}

\date{2022}

\pubyear{2022}

\begin{document}
\label{firstpage}
\pagerange{\pageref{firstpage}--\pageref{lastpage}}
\maketitle

\begin{abstract}
Using new high-resolution data of CO~(2--1), H$\upalpha$ and H$\upbeta$
obtained with the Northern Extended
Millimeter Array (NOEMA) and the Multi-Unit Spectroscopic Explorer (MUSE)
at the Very Large Telescope, we have performed a Toomre~$Q$ disc stability
analysis and studied star formation, gas depletion times and
other environmental parameters on sub-kpc scales within the $z\sim0$ galaxy
SDSS J125013.84+073444.5 (LARS~8).
The galaxy hosts a massive, clumpy disc and is a proto-typical analogue
of main-sequence galaxies at $z\sim1-2$.
We show that the massive (molecular) clumps in LARS~8 are the result of
an extremely
gravitationally unstable
gas disc,
with large scale instabilities found across the whole extent of the
rotating disc, with only the innermost 500\,pc being stabilized by its
bulgelike structure.
%With the extinction-corrected MUSE H$\upalpha$ map, we further established
%the resolved star-formation law (Kennicutt-Schmidt relation) for LARS 8 and
The radial profiles further reveal that
-- contrary to typical disc galaxies -- the molecular gas
depletion time decreases from more than 1\,Gyr in the center to less than
$\sim$100\,Myr in the outskirts of the disc, supporting the findings of a
Toomre-unstable disc.
We further identified and analysed 12 individual massive molecular clumps.
They are virialized and follow the mass-size relation, indicating
that on local (cloud/clump) scales the stars form with efficiencies comparable
to those in Milky Way clouds. The observed high
star formation rate must thus be the result of triggering of cloud/clump
formation over large scales due to disc instability.
Our study provides evidence that ``in-situ'' massive clump formation (as also observed at
high redshifts) is very efficiently induced by large-scale instabilities.
\end{abstract}

\begin{keywords}
galaxies: starburst -- galaxies: star formation -- galaxies: ISM -- galaxies: kinematics and dynamics -- techniques: interferometric -- techniques: imaging spectroscopy
\end{keywords}

%%%%%%%%%%%%%%%%%%%%%%%%%%%%%%%%%%%%%%%%%%%%%%%%%%

%%%%%%%%%%%%%%%%% BODY OF PAPER %%%%%%%%%%%%%%%%%%

%%%%%%%%%%%%%%%%%%%%%%%%%%%%%%%%%%%%%%%%%%%%%%%%%%%%%%%%%%%%%%%%%%%%
%%%%%%%%%%%%%%%%%%%%%%%%%%%%%%%%%%%%%%%%%%%%%%%%%%%%%%%%%%%%%%%%%%%%
%%%%%%%%%%%%%%%%%%%%%%%%%%%%%%%%%%%%%%%%%%%%%%%%%%%%%%%%%%%%%%%%%%%%
%%%%%%%%%%%%%%%%%%%%%%%%%%%%%%%%%%%%%%%%%%%%%%%%%%%%%%%%%%%%%%%%%%%%
\section{Introduction}
Several studies based on deep field observations
have revealed that at redshifts $\sim$3 galaxies with
total (gas+stars) masses similar to the Milky Way ($\sim$\,$10^{11}$\,M$_{\odot}$) are already
in place \citep{Dessauges-Zavadsky2017,Elbaz2018,Tacconi2018,Cassata2020}.
Since the Universe was then only $\sim$2\,Gyr old, these
massive objects must have formed within a very short time,
thus requiring very high star formation rates (SFRs)
compared to
the $z\sim0$ Universe.
In recent years, observations of galaxies at redshifts between 0 and 4 have shown that the level of star formation
is mainly dictated by stellar mass and regulated by secular processes \citep{Popesso2019}.
This is
manifested in a
tight relation between stellar mass and SFR, the so called
\textit{main sequence of star forming galaxies}
\citep{Brinchmann2004,Noeske2007,Daddi2007,Elbaz2007,Peng2010,Wuyts2011,Whitaker2012,Whitaker2014,Tomczak2016}.
While the slope of the relation does not vary with redshift, its intercept shifts towards higher SFRs
with increasing lookback time,
see for example \cite{Wuyts2011,Rodighiero2011}.
As shown by several studies \citep{Tacconi2013,Genzel2015,Scoville2017,Wiklind2019}, the evolution of the main
sequence is driven by an increase of the (molecular) gas fraction. As a result the gas depletion time, 
defined as the
inverse
of the SFR per unit molecular gas mass, remains roughly constant even out to $z\sim$4. The
high SFRs observed on the main sequence at high-$z$ are thus mainly driven by increasing gas fractions.
Although the strong evolution of the SFR with time (redshift) is relatively well constrained \citep{Madau2014},
the underlying physical mechanisms that drive star formation in
gas-rich discs is still a matter of debate.

%For example, it was realized that the gas consumption time scales are too short compared to the age of the Universe.
%Thus, galaxies must accrete gas from their circumgalactic medium in order to sustain star formation at the observed high rates.
%A model was put forward, the so called \textit{regulator} \citep{Lilly2013} or \textit{bathtub} model \citep{Bouche2010},
%in which the accretion rate and the SFR are in equilibrium.
%
%More recently, Genzel+ (2011, 2014) have observed clumpy disc galaxies at redshifts z$\sim$2--3
%and estimated the two-dimensional distribution of the Toomre $Q$ parameter
%of the perturbed disc \textit{from the gas component only}.
%They found $Q<1$ in the extended disc regions, but $Q>1$ in inner regions that are bulge dominated.
%This seems to be consistent with standard linear Toomre instability in an outer ring and morphological
%quenching inside (Martig+ 2009).

Beside their large gas fractions, high-$z$ main-sequence galaxies
are
observed to
have higher gas velocity dispersions \citep{FoersterSchreiber2009,Lehnert2009,Swinbank2012,Wisnioski2015}
compared to local spirals.
Additionally,
their morphologies show extremely massive \textit{clumps}, exhibiting considerable fractions of the total mass.
Initially, this was interpreted
in the context of ``bottom up'' structure formation as an ongoing process of merging.
However, with the advent of near-infrared integral field spectroscopy,
in some of the clumpy galaxies, disc structures were found that are characterized by significant
rotation \citep{Genzel2006,Genzel2008}, a sign for associated structures rather than mergers.
However, some fraction of them may still be ongoing mergers \citep{Weiner2006,FoersterSchreiber2009,Puech2010,Rodrigues2017}.
The observed discs host giant clumps with masses of $M_\mathrm{cl}\lesssim10^9$\,M$_{\odot}$.
It has been proposed that such clumps result from the fragmentation of massive gas discs driven
by gravitational instability \citep{Agertz2009b,Dekel2009,Bournaud2012,Romeo2014}.
Thus, the mode of star formation in gas-rich systems seems fundamentally different compared
to the star formation within spiral arms as found in most galaxies at $z\sim0$.

Much numerical work has been undertaken in the last few years to study gravitational fragmentation scenarios.
But in early simulations,
inefficient thermal feedback of supernovae resulted in overcooling, which then
enhanced disc instability and star formation and led to an overproduction of giant clumps \citep{Ceverino2009,Agertz2009b}.
More recently, various works within cosmological simulations and simulations of isolated disc galaxies
were again focusing on disc fragmentation at high-$z$,
but including novel feedback recipes which systematically led to
less fragmentation even in massive gas-rich discs and generally
lower clump masses in the range $10^7$--$10^8$\,M$_{\odot}$
\citep{Tamburello2015,Behrendt2015,Moody2014,Mandelker2017,Oklopcic2017}.
It was further proposed that some of the most massive observed star forming clumps
might not be the result of ``in-situ'' disc fragmentation, but rather they could be
accreted cores of massive satellite galaxies \citep{Mandelker2017,Oklopcic2017}.
Thus, to date the question of whether massive clumps are a result of ``in-situ'' disc fragmentation
or the product of accreted cores of massive satellite galaxies remains unanswered and
needs iterations on both ends, theory and observations. The observational difficulty is
that at high-$z$ the spatial resolution is often too coarse to
constrain relevant physical parameters (turbulence, density, timescales related to star formation).
Only gravitational lensing may help to reveal details about the clumpy discs
at $z\sim1-3$ \citep{Dessauges-Zavadsky2019}.

Despite the difficulties, recent observations support
galaxy-wide disc instabilities as a cause of the clumpy nature, e.g.
\cite{Dessauges-Zavadsky2018} showed that the clump mass function
at $z\sim$1--3 follows a power-law consistent with turbulence being the driving
mechanism. Moreover, the DYNAMO survey \citep{Fisher2014},
targeting extremely rare \textit{local} clumpy gas-rich disc galaxies as a proxy
for high-$z$ galaxies, revealed clump properties that favour
clump formation induced by galaxy-wide disc instabilities
\citep{Fisher2017,White2017,Fisher2019}.

\begin{figure}
	\includegraphics[width=\columnwidth]{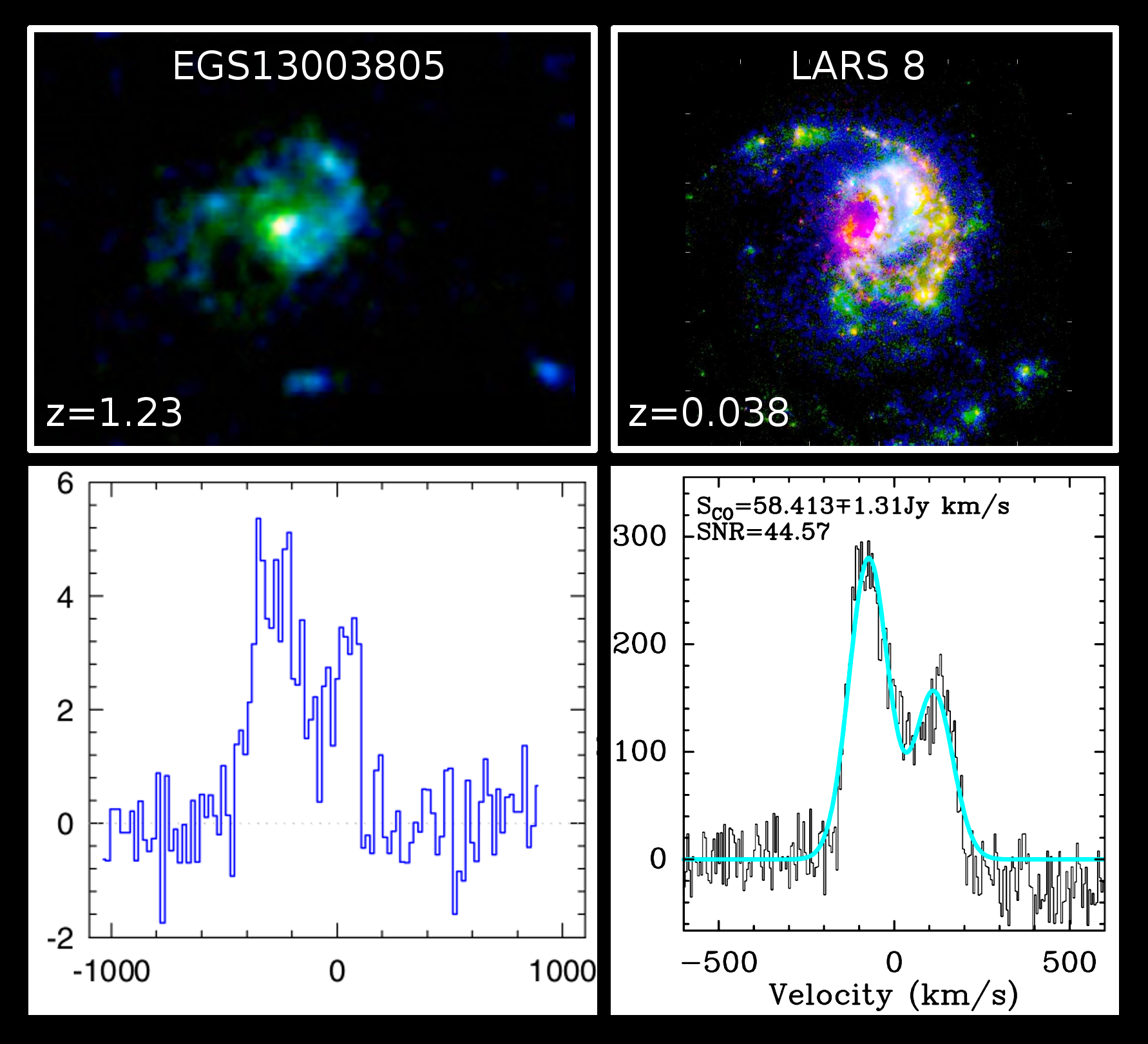}
    \caption{High-$z$ target from the PHIBBS survey \citep{Tacconi2013} versus LARS 8 \citep{Puschnig2020}.
    The optical morphologies (\textit{top panel}) as well as the CO line emission (\textit{bottom panel}) are remarkably similar.}
    \label{fig:lars08_phibbs}
\end{figure}

In this paper, we present new highly-resolved NOEMA CO (2--1) and MUSE H$\upalpha$
observations of LARS~8, a clumpy $z\sim0$
galaxy drawn from the \textit{Lyman Alpha Reference Sample} \citep{Oestlin2014,Hayes2014}.
Given the basic properties of the galaxy (see Table \ref{tab:lars8props}) with
a stellar mass of $\sim$10$^{11}$ M$_\odot$ and a SFR of $\sim$30 M$_\odot$\,yr$^{-1}$,
LARS~8 resembles main-sequence galaxies at high-$z$.
The LARS galaxy is also known to be rotationally supported \citep{Herenz2016,Puschnig2020},
just like face-on disc galaxies typically observed at high-$z$ (compare Figure \ref{fig:lars08_phibbs}).
\cite{Herenz2016} and \cite{Micheva2018} further revealed the existence of
shells at large galactocentric radii, caused by a merger event that LARS~8
must have undergone recently. Using deep high-resolution 21\,cm observations, \cite{LeReste2022}
found a large neutral gas reservoir westwards of the optical galaxy disk.
The galaxy is known to have a relatively high gas fraction of 27 percent,
a gas depletion time of $\sim$1.2\,Gyr \citep{Puschnig2020}
and a clumpy morphology.
These properties make LARS~8 an ideal laboratory to study clump formation in
a gas-rich disc galaxy.

The paper is organised as follows. In Section \ref{sec:observations} we inform
about the spectroscopic observations our results are based on. The methods and tools we
use to convert the observables into physical parameters (e.g. star formation rates, mass surface densities, dynamical parameters)
are outlined in Section \ref{sec:methods}. The results are presented in Section \ref{sec:results}
and subsequently discussed and compared to related works in \ref{sec:discussion}. Section
\ref{sec:summary} concludes the paper with a summary.

Throughout the paper, we adopt a cosmology with $H_\mathrm{0}$=70, $\Omega_\mathrm{M}$=0.3 and $\Omega_\mathrm{vac}$=0.7.

\begin{table}
\caption{Properties of LARS 8 (see Tables 1, 9 and 10 of Puschnig et al. 2020). $\dagger$ The metallicity $Z$ was derived by
\"Ostlin et al. (2014) using the $R_\mathrm{23}$-$P$ relation and is given in units of $\mathrm{12+log(O/H)}$. $\ddag$ The blue-band diameter of the 25\,mag\,arcsec$^{-2}$ isophote was derived from SDSS g-band observations using an SQL query on SDSS DR7. Since the SDSS g-band isophote
is typically $\sim$1.3 times larger than those measured in the Johnson B band, the
g-band diameter was divided by that factor.}\label{tab:lars8props}
\begin{tabular}{ccccccc}
\hline \hline
& $D_\mathrm{L}$     & log $M_*$       & $Z^\dagger$       & SFR                     & $D_\mathrm{25}^{\ddag}$ & $f_\mathrm{gas}$ \\
& [Mpc]              & [M$_\odot$]     &                   & [M$_\odot$ yr$^{-1}$]   & ["]                     & [\%] \\
& 167.5$\pm$12       & 10.97$\pm$0.10  & 8.51              & 30$\pm$8                & 30.8                    & 27$\pm$16 \\
\end{tabular}
\end{table}

%%%%%%%%%%%%%%%%%%%%%%%%%%%%%%%%%%%%%%%%%%%%%%%%%%%%%%%%%%%%%%%%%%%%
%%%%%%%%%%%%%%%%%%%%%%%%%%%%%%%%%%%%%%%%%%%%%%%%%%%%%%%%%%%%%%%%%%%%
%%%%%%%%%%%%%%%%%%%%%%%%%%%%%%%%%%%%%%%%%%%%%%%%%%%%%%%%%%%%%%%%%%%%
%%%%%%%%%%%%%%%%%%%%%%%%%%%%%%%%%%%%%%%%%%%%%%%%%%%%%%%%%%%%%%%%%%%%
\section{Observations and data reduction}\label{sec:observations}

%%%%%%%%%%%%%%%%%%%%%%%%%%%%%%%%%%%%%%%%%%%%%%%%%%%%%%%%%%%%%%%%%%%%
%%%%%%%%%%%%%%%%%%%%%%%%%%%%%%%%%%%%%%%%%%%%%%%%%%%%%%%%%%%%%%%%%%%%
\subsection{NOEMA CO (2--1) cube}
We observed LARS~8 in a single pointing under programs W16BS
and E16AG with the IRAM Northern Extended Millimeter Array (NOEMA) using eight antennas in 
configurations A and D, providing maximum baselines of $\sim$760\,m and $\sim$180\,m respectively.
The target line, CO~(2--1), was observed with the WideX correlator
(bandwidth $\sim$3.6\,GHz) using a tuning frequency of 222.044\,GHz, corresponding to
the systemic velocity derived from H~I observations
\citep{Pardy2014}.
We further performed \textit{on-the-fly}
mapping of LARS~8 with the IRAM 30m telescope under programs 064-15 and 178-15,
allowing us to include short spacing visibility data. 

Extended array observations were executed on December 15, 2016 for a total
on-source time of 5.2 hours under good weather conditions
with a precipitable water vapour (PWV) of $\sim$1.8\,mm. Compact array observations were executed on
three days during May 2017 for a total on-source time
of 6.3 hours under average weather conditions with a PWV of $\sim$2--3\,mm.

The absolute flux scale of the configuration A data was calibrated on LKHA101 using 
a model flux of 0.54\,Jy. The sources 1222+216 and 3C273 were used as phase and amplitude calibrators.
Average polarization mode was chosen for the amplitude calibration, because the
signal was found to be polarized. 3C84 was used as bandpass calibrator. 

The absolute flux scale of the configuration D data obtained on May 8, 2017
was calibrated on MWC349 with a model flux of 1.87\,Jy. For the observations executed on
May 2 and May 3, MWC349 data was not available and 3C273 was used instead, assuming
a model flux of 7.65\,Jy, as measured on May 8. 1236+077 and 3C273 were used for
phase and amplitude calibration, whereas
3C273 was also used as bandpass calibrator.

All observations were calibrated using the IRAM reduction pipeline
\texttt{GILDAS/CLIC}\footnote{\url{http://www.iram.fr/IRAMFR/GILDAS/}}.
Data flagging was performed manually taking into account tracking
errors, pointing and focus offsets as well as quality assessment
through outlier rejection in time versus amplitude and phase plots
as well as large phase discrepancies between the two polarizations.
We remark that
for the configuration A observations, for one of the antennas tracking
errors of more than 4" were reported. All baselines including this antenna were thus
rejected. 

Merging of the calibrated visibilities and short spacing correction
were performed in the \texttt{GILDAS/mapping} environment, which was
also used for imaging. Robust weighting of 0.5 was found to lead to
a good compromise between sidelobe suppression and spatial resolution,
both of which are important for our science case. Cleaning was done
using the H\"ogbom algorithm \citep{Hogbom1974} and a central circular support of 6" diameter.
The final clean cube has an r.m.s. noise of 0.8\,mJy/beam
at a velocity resolution of 10\,km/s. The synthetic beam size
is 0.61"x0.37" with a position angle of 36\textdegree. 

As LARS~8 is substantially extended compared to NOEMA's $\sim$23" field of view
at the observed frequency, a primary beam correction was finally
performed using the \texttt{PRIMARY} task
within the \texttt{GILDAS/mapping} environment.

%%%%%%%%%%%%%%%%%%%%%%%%%%%%%%%%%%%%%%%%%%%%%%%%%%%%%%%%%%%%%%%%%%%%
%%%%%%%%%%%%%%%%%%%%%%%%%%%%%%%%%%%%%%%%%%%%%%%%%%%%%%%%%%%%%%%%%%%%
\subsection{MUSE observations and data reduction}
We observed LARS~8 with the Multi-Unit Spectroscopic Explorer (MUSE; \citealt{Bacon2010}) integral field spectrograph, mounted at Unit Telescope 4 of the Very Large Telescope (VLT).  Spectra were obtained on the night of 18 May 2018 under conditions of new moon, airmass lower than 1.2, and with a V-band seeing of 0\farcs8.  We obtained four observations of the main target, each rotated by 90 degrees compared to the previous to minimize fixed pattern noise from the image slicers, using integration times of 650 seconds.  Because LARS~8 occupies a major fraction of the MUSE field-of-view, we also obtained a separate sky frame from an adjacent empty pointing using an integration time of 120 seconds.
Data were reduced using Version 2.6 of the ESO pipeline, using standard methods and paying special attention to the removal of low surface brightness emission in strong nebular lines. 

%Subtraction of the stellar continuum was performed with... I guess Eric writes the remainder...

%%%%%%%%%%%%%%%%%%%%%%%%%%%%%%%%%%%%%%%%%%%%%%%%%%%%%%%%%%%%%%%%%%%%
%%%%%%%%%%%%%%%%%%%%%%%%%%%%%%%%%%%%%%%%%%%%%%%%%%%%%%%%%%%%%%%%%%%%
%%%%%%%%%%%%%%%%%%%%%%%%%%%%%%%%%%%%%%%%%%%%%%%%%%%%%%%%%%%%%%%%%%%%
%%%%%%%%%%%%%%%%%%%%%%%%%%%%%%%%%%%%%%%%%%%%%%%%%%%%%%%%%%%%%%%%%%%%
\section{Methods}\label{sec:methods}
Here we briefly describe the routines and tools
that we used to obtain physical parameters from the observations.
In subsections \ref{sec:method_observables_start}--\ref{sec:method_observables_end}
we explain how the observed
data cubes are prepared for further scientific analysis, i.e.
the convolution to a common resolution and the derivation of
moment maps.

Subsections \ref{sec:method_physpar_start}--\ref{sec:method_physpar_end}
summarize the assumptions and constraints used to convert the
observables into physical quantities such as star formation rates,
stellar and gas surface densities.

In \ref{sec:rotcurve} and \ref{sec:stellar_vdisp} the routines
for the characterisation of the dynamics of the galaxy's gaseous and
stellar components are presented. These dynamical quantities finally
allow to constrain the gravitational instability via the Toomre parameter (see Section \ref{sec:toomreQ}).

We conclude in \ref{sec:clump_identification} and
\ref{sec:pde} with a brief description of 
how individual molecular gas clouds are identified in the NOEMA cube
and how we use previously derived physical quantities to estimate
the dynamic equilibrium pressure.

%%%%%%%%%%%%%%%%%%%%%%%%%%%%%%%%%%%%%%%%%%%%%%%%%%%%%%%%%%%%%%%%%%%%
%%%%%%%%%%%%%%%%%%%%%%%%%%%%%%%%%%%%%%%%%%%%%%%%%%%%%%%%%%%%%%%%%%%%
\subsection{Convolution of the NOEMA data cube to a common resolution}\label{sec:method_observables_start}
Given the slightly lower spatial resolution of the optical data cube compared to our radio data,
we convolve the latter to match the resolution of the MUSE observations.
To do so, we first deconvolve the elliptical NOEMA beam from the circularized
target beam (based on MUSE cube). The resulting convolution kernel is then applied onto the 3D
NOEMA cube (plane-by-plane) using the \texttt{scipy.signal.convolve} package \citep{SciPy2020}.
The circularized
synthetic beam size of the matched-resolution NOEMA CO (2--1) data cube is 0.78".
Note that throughout the paper we make use of the native resolution CO (2--1) data whenever
possible (clump identification, Toomre disc stability analysis). Only plots
that include both star formation rates (from H$\upalpha$) and properties
derived from the CO observations are based on the matched-resolution data.

%%%%%%%%%%%%%%%%%%%%%%%%%%%%%%%%%%%%%%%%%%%%%%%%%%%%%%%%%%%%%%%%%%%%
%%%%%%%%%%%%%%%%%%%%%%%%%%%%%%%%%%%%%%%%%%%%%%%%%%%%%%%%%%%%%%%%%%%%
\subsection{MUSE line extraction of H$\alpha$, H$\beta$ and continuum subtraction}
From the reduced MUSE cube, we first extract a fixed spectral range around the
observed H$\upalpha$ and H$\upbeta$ lines, using $z$=0.0382531 as the redshift and
an extraction window of $\pm$420\,km/s, centered on the systemic line-center (corresponding to $z$). 
We ensured that the [N~II] lines are
outside the extracted line window of H$\upalpha$.
In order to define the continuum level at each line,
individual spectral windows were defined
blueward and redward
of each emission line,
after manual inspection of the spectral cube.
For H$\upalpha$, suitable windows
were found between -2500 and -1500\,km/s and from 3000 to 4000\,km/s.
H$\upbeta$ continuum levels were evaluated between -4000 and -2000\,km/s as well as
within the range of 2000 and 4000\,km/s.
The continuum correction for each line was then performed via subtraction of a linear fit,
obtained from regression (using the python \texttt{lmfit} package)
of the flux within the given velocity intervals.

%%%%%%%%%%%%%%%%%%%%%%%%%%%%%%%%%%%%%%%%%%%%%%%%%%%%%%%%%%%%%%%%%%%%
%%%%%%%%%%%%%%%%%%%%%%%%%%%%%%%%%%%%%%%%%%%%%%%%%%%%%%%%%%%%%%%%%%%%
\subsection{Moment maps of CO (2--1) and optical emission lines}\label{sec:method_observables_end}
We generate moment zero maps of CO~(2--1), H$\upalpha$ and H$\upbeta$ via summation of the flux
in masked channels, using the approach of ``dilated masking''.
In the NOEMA cube,
peak channels were identified that have a more than 4-sigma strong signal in
at least three adjacent channels. The mask was then expanded in velocity space
as long as the flux in two adjacent channels was above a 2-sigma limit.
Additionally, we only allow connected spatial regions that cover
at least the size of the synthetic beam of our observations.
Moment maps of H$\upalpha$ and H$\upbeta$ were created in a very similar manner,
i.e. we identified channels with 4-sigma peaks and subsequently grow the mask
down to a level of 2-sigma. However, given the lower spectral resolution of the
MUSE data cubes, we allow to mask even single channels in velocity space rather than a number of adjacent ones.

\begin{figure*}
	\includegraphics[width=\textwidth]{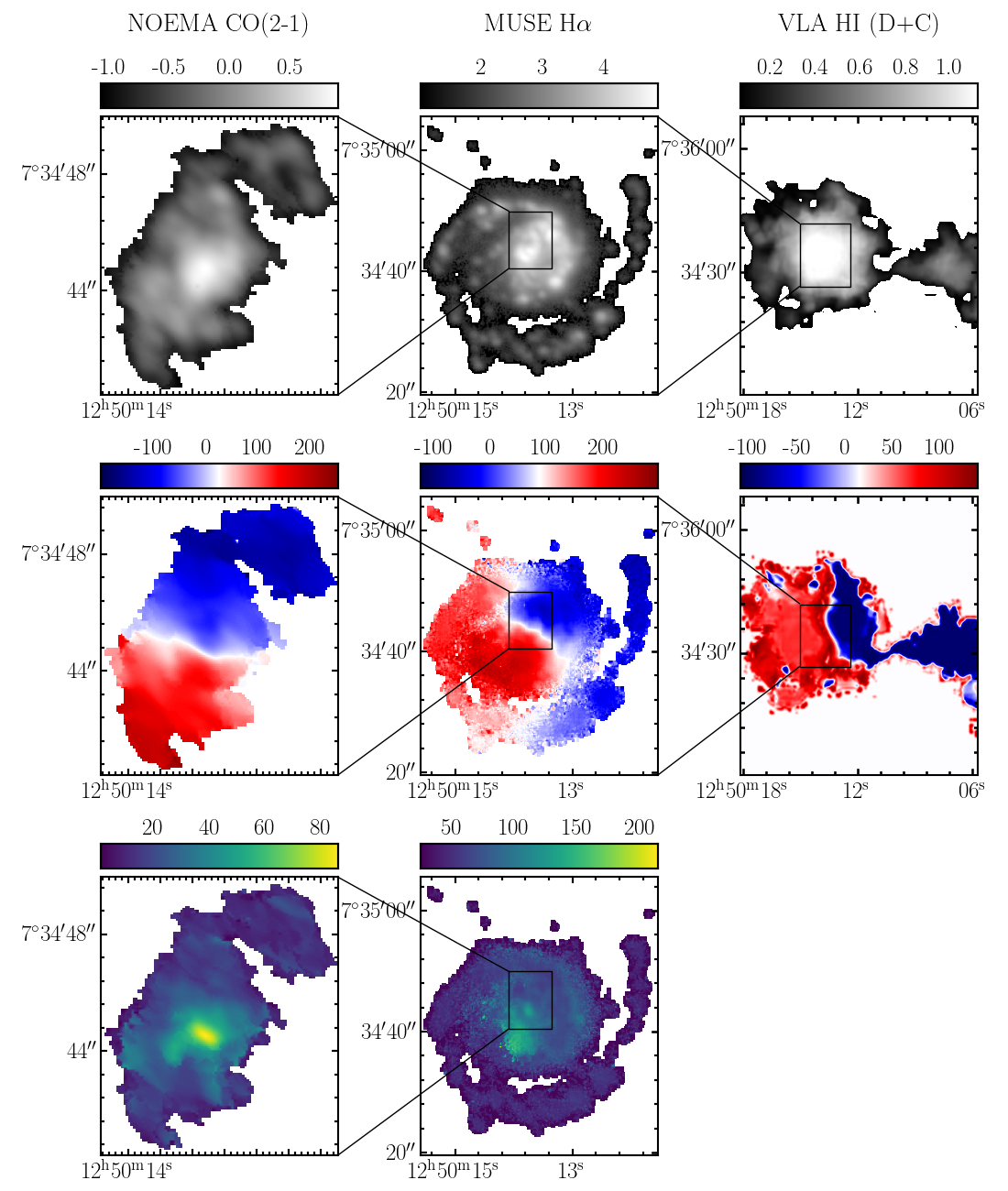}
    \caption{Maps derived from NOEMA CO (\textit{left column}), MUSE H$\upalpha$ (\textit{middle column}) and
    VLA HI 21cm (\textit{right column}; \citealt{LeReste2022}) data cubes. The top row moment-0 maps are
    given in units of Jy\,km\,s$^{-1}$\,beam$^{-1}$ on a logarithmic scale for CO, 10$^{-20}$ erg\,s$^{-1}$\,cm$^{-2}$ on a logarithmic scale for H$\upalpha$
    and 10$^{21}$ cm$^{-2}$ on a linear scale for HI. The moment-1 and moment-2 maps in the middle and bottom rows are given in km\,s$^{-1}$.}
    \label{fig:moments}
\end{figure*}

First and second moment maps were created using the same masks, with
moment one being the intensity-weighted mean velocity found
under the masked channels and moment two being the
intensity-weighted r.m.s. velocity scatter.
Moment maps are shown in Figures \ref{fig:moments}.

The uncertainties of our moment maps are calculated via Gaussian error propagation
using the r.m.s. outside the line masks as an estimate for the uncertainty of
each masked channel.

%%%%%%%%%%%%%%%%%%%%%%%%%%%%%%%%%%%%%%%%%%%%%%%%%%%%%%%%%%%%%%%%%%%%
%%%%%%%%%%%%%%%%%%%%%%%%%%%%%%%%%%%%%%%%%%%%%%%%%%%%%%%%%%%%%%%%%%%%
\subsection{Star formation rates from MUSE H$\alpha$}\label{sec:method_physpar_start}
In order to obtain the intrinsic, extinction-corrected H$\upalpha$ flux, we
calculate the dust attenuation from the Balmer decrement using the
\cite{Cardelli1989} attenuation law and assume case~B recombination
at 10$^4$\,K
and an intrinsic, theoretical H$\upalpha$/H$\upbeta$ ratio of 2.86.
The average extinction E(B-V) within an aperture of 5 arcsec radius enclosing
the center of the galaxy -- and thus covering the main part of the NOEMA field-of-view --
is 0.7\,mag with a maximum value of 1.2\,mag in the central pixel and
values as low as 0.2\,mag in the outer region.

We first convert the extinction-corrected H$\upalpha$ flux from erg/s/cm$^2$
into the corresponding luminosity ($L_\mathrm{H\upalpha}$) in erg/s using a luminosity-distance of
167.5\,Mpc. The star formation rates (SFRs) in units of M$_\odot$~yr$^{-1}$ per pixel are then calculated using the
calibration of \cite{Calzetti2012}: SFR=5.5~10$^{-42}$~$L_\mathrm{H\upalpha}$. These
SFRs are then converted into surface densities in units of M$_\odot$~yr$^{-1}$~kpc$^{-2}$
($\Sigma_\mathrm{SFR}$) taking into account the cosine correction factor
($\mathrm{cos}\ i$)
for the galaxy inclination $i$ of 50$^\circ$ that we found from the
rotation curve (see Section \ref{sec:rotcurve}).

%%%%%%%%%%%%%%%%%%%%%%%%%%%%%%%%%%%%%%%%%%%%%%%%%%%%%%%%%%%%%%%%%%%%
%%%%%%%%%%%%%%%%%%%%%%%%%%%%%%%%%%%%%%%%%%%%%%%%%%%%%%%%%%%%%%%%%%%%
\subsection{Stellar surface density}
A stellar mass map of the galaxy is constructed by performing a pixel spectral energy distribution fit using the 
HST FUV and optical broad band data from the LARS project. The fitting code ``the Ly$\upalpha$ eXtraction Software''
(\citealt{Oestlin2014}, Melinder et al. in preparation) uses two single stellar populations with four free parameters: stellar mass 
for the two components, stellar age, and stellar extinction (only one of the populations have a varying age and extinction, 
the other one is kept at an age of 10\,Gyrs and an E(B-V)$_s$ of 0). The fit is performed for each pixel (or spatial bin) 
to produce maps of stellar continuum fluxes, mass, age, and extinction.
The uncertainties on the stellar masses are estimated within the code using Monte Carlo simulations, in which random noise
(corresponding to the r.m.s. in each pixel after drizzling) is added to the originally measured value. The final uncertainty is
then the standard deviation obtained from the measurements in all Monte Carlo simulations.
For details on the code and the data used for 
LARS~8 we refer the reader to \cite{Oestlin2014}.
To find the stellar surface density radial profile we co-add the 
stellar mass maps of the two components and measure the mean mass surface density in elliptical annuli that exactly
match those used for the NOEMA and MUSE data. Finally, the derived mean surface densities ($\Sigma_*$) were corrected for
inclination using the same quantities as for $\Sigma_\mathrm{SFR}$.
The scale length of the stellar disc, $l_*$, was derived from the stellar mass map by 
fitting an exponential function to the inclination corrected mass profile.

%%%%%%%%%%%%%%%%%%%%%%%%%%%%%%%%%%%%%%%%%%%%%%%%%%%%%%%%%%%%%%%%%%%%
%%%%%%%%%%%%%%%%%%%%%%%%%%%%%%%%%%%%%%%%%%%%%%%%%%%%%%%%%%%%%%%%%%%%
\subsection{Molecular gas surface density, depletion time and gas fraction}\label{sec:method_physpar_end}
We convert our measured CO (2--1) fluxes $S_\mathrm{CO}$
(in units of Jy~km~s$^{-1}$~beam$^{-1}$) to CO luminosities using the definition of $L'_\mathrm{CO}$ by \cite{Solomon2005}:
\begin{equation}\label{eqn:LPCO}
L'_\mathrm{CO(2-1)} = 3.25\ 10^7\ S_\mathrm{CO(2-1)}\ \nu_\mathrm{obs}^{-2}\ D_\mathrm{L}^2\ (1+z)^{-3}
\end{equation}
$L'_\mathrm{CO(2-1)}$ is then given in K\,km\,s$^{-1}$\,pc$^2$, $z$ is the redshift, $D_\mathrm{L}$ the luminosity distance in Mpc and
$\nu_\mathrm{obs}$ the observed frequency in GHz. For the conversion from $L'_\mathrm{CO(2-1)}$ to
molecular gas masses we first need to down-convert to the luminosity of the J=1--0 line ($L'_\mathrm{CO(1-0)}$),
for which we assume a line ratio CO(2--1)/(1--0) of 0.7,
which is typically observed in several types of galaxies \citep{Saintonge2017,denBrok2021}.
Subsequent multiplication with the conversion
factor $\alpha_\mathrm{CO}$ finally leads to the molecular gas masses ($M_\mathrm{H_2}$).
Here we use $\alpha_\mathrm{CO}$ = 4.5 that we derived previously
using a metallicity-dependent approach \citep{Puschnig2020}.
This value is similar to $\alpha_\mathrm{CO}$ in the Milky Way \citep{Bolatto2013}.
We stress that our choice of $\alpha_\mathrm{CO}$ is based on a
galaxy-wide average. A lower conversion factor might be applicable in the
center of the galaxy due to lower CO optical depths driven by a
large velocity dispersion. However, to date, no data (e.g. $^{13}$CO) is available to
assess any radial trend of the conversion factor in LARS~8.
Again, the final gas mass surface density map ($\Sigma_\mathrm{{H_2}}$) was corrected for the inclination of the galaxy.
The molecular gas depletion time \tdep\ and the gas fraction $f_\mathrm{gas}$
were calculated in the following way:
\begin{equation}
    \tau_\mathrm{{depl}}\ =\ \frac{M_\mathrm{{H_2}}}{SFR}
\end{equation}
\begin{equation}
    f_\mathrm{{gas}}\ =\ \frac{M_\mathrm{{H_2}}}{M_\mathrm{{H_2}}\ +\ M_*}
\end{equation}

%%%%%%%%%%%%%%%%%%%%%%%%%%%%%%%%%%%%%%%%%%%%%%%%%%%%%%%%%%%%%%%%%%%%
%%%%%%%%%%%%%%%%%%%%%%%%%%%%%%%%%%%%%%%%%%%%%%%%%%%%%%%%%%%%%%%%%%%%
\subsection{Molecular gas rotation curve analysis}\label{sec:rotcurve}
We use \texttt{$^{3D}$BAROLO} \citep{3DBAROLO} to
derive the galaxy rotation curve from the NOEMA CO~(2--1)
data cube. The software iteratively fits 3D tilted-ring models to the cube
and solves in each ring for inclination, position angle (PA), rotation velocity
and velocity dispersion. We ran the software several times to experiment
with input parameters such as pixel coordinates of the kinematic center,
fixating systemic velocity, inclination and/or PA. Despite the fact
that the algorithm robustly constrained inclination and PA we ultimately
decided to fix the two parameters (for each ring) to 50 and 160 degrees respectively,
while leaving the position of the kinematic center as free parameter.
The free parameters derived for each 0.61" wide ring (i.e.\ the major axis of the beam)
are thus the rotation velocity, the velocity dispersion ($\sigma_{g}$) and the coordinates of the kinematic center.
A summary of the results is shown in Table \ref{tab:barolo}.
The position-velocity diagram and the smoothed and interpolated rotation curve
obtained using these constraints are shown in
Figures \ref{fig:3dbarolo_pv} and \ref{fig:rotcurve}. A comparison between observed and
modeled quantities is found in Appendix \ref{fig:3dbarolo_obs_model}.

\begin{figure}
	\includegraphics[width=\columnwidth]{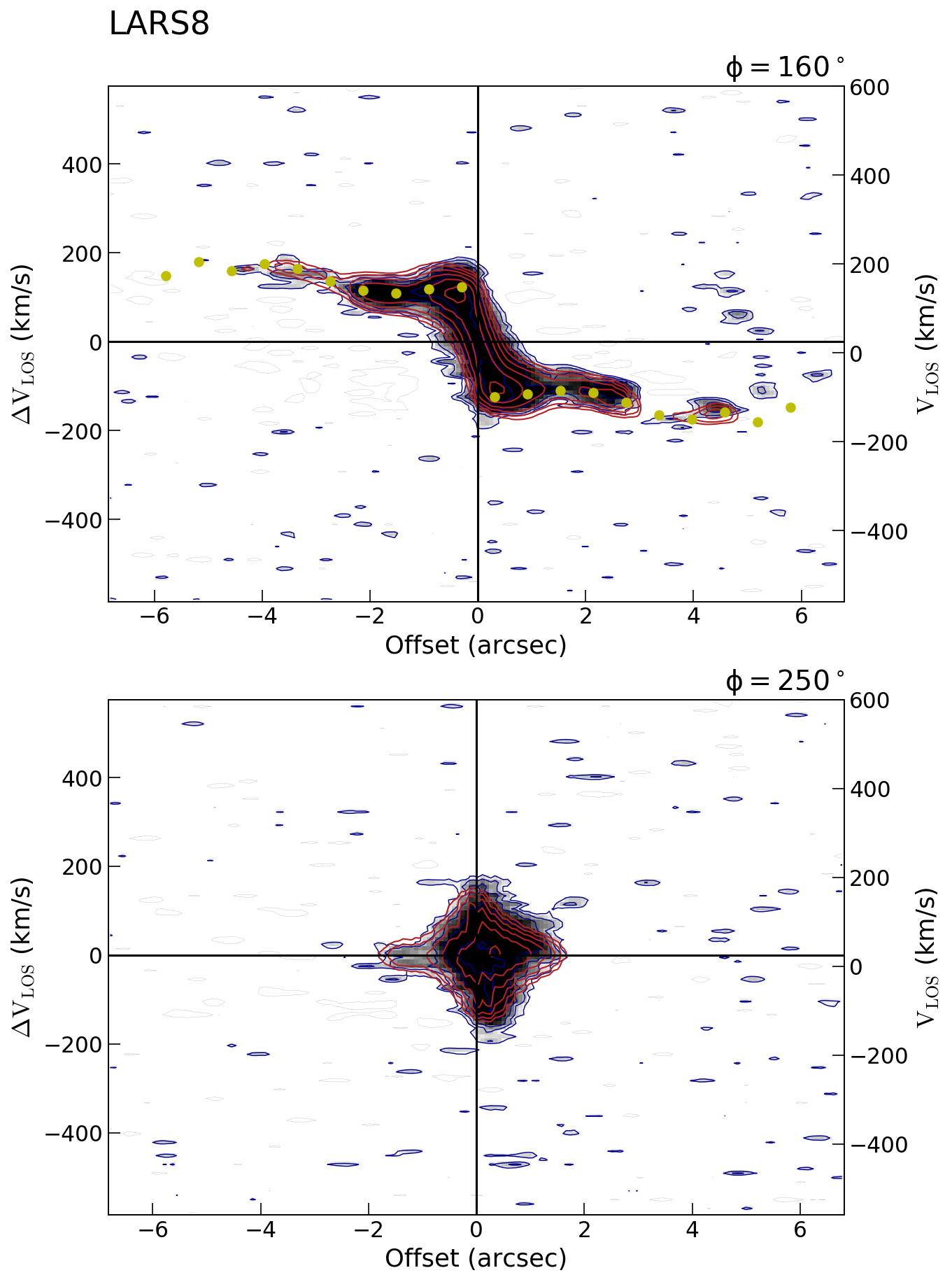}
    \caption{NOEMA CO (2--1) position-velocity diagram along the major axis (\textit{top panel}) and minor axis (\textit{bottom panel}) of LARS 8.
    The yellow-brown points indicate the radial bins of the derived rotation curve.}
    \label{fig:3dbarolo_pv}
\end{figure}

\begin{figure}
	\includegraphics[width=\columnwidth]{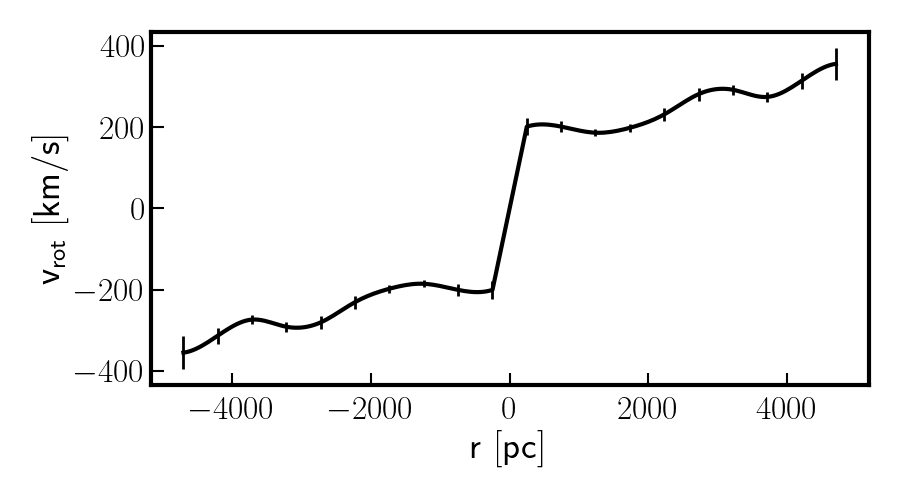}
    \caption{Smoothed and interpolated rotation curve (before inclination correction) derived for LARS 8 from the NOEMA CO (2--1) cube.}
    \label{fig:rotcurve}
\end{figure}

\begin{table}
\caption{Kinematic properties for elliptical rings.}\label{tab:barolo}
\begin{tabular}{ccccc}
\hline \hline
\begin{tabular}[c]{@{}l@{}}Radius\\ {[}"{]}\end{tabular} & \begin{tabular}[c]{@{}l@{}}v$_\mathrm{rot}$\\ {[}km/s{]}\end{tabular} & \begin{tabular}[c]{@{}l@{}}$\sigma_g$\\ {[}km/s{]}\end{tabular} & \begin{tabular}[c]{@{}l@{}}i\\ {[}°{]}\end{tabular} & \begin{tabular}[c]{@{}l@{}}PA\\ {[}°{]}\end{tabular}  \\
0.305              & 154$\pm$17                  & 11$\pm$4                  & 50        & 160          \\
0.915              & 154$\pm$11                  & 28$\pm$7                  & 50        & 160                \\
1.525              & 142$\pm$7                  & 20$\pm$4                  & 50        & 160               \\
2.135              & 151$\pm$7                  & 17$\pm$4                  & 50        & 160             \\
2.745              & 176$\pm$12                  & 17$\pm$6                  & 50        & 160             \\
3.355              & 215$\pm$12                  & 11$\pm$7                   & 50        & 160             \\
3.965              & 223$\pm$10                  & 9$\pm$5                   & 50        & 160             \\
4.575              & 210$\pm$9                  & 6$\pm$5                   & 50        & 160            \\
5.185              & 240$\pm$15                  & 8$\pm$4                   & 50        & 160               \\
5.795              & 272$\pm$30                  & 7$\pm$4                   & 50        & 160           \\
\end{tabular}
\end{table}

%%%%%%%%%%%%%%%%%%%%%%%%%%%%%%%%%%%%%%%%%%%%%%%%%%%%%%%%%%%%%%%%%%%%
%%%%%%%%%%%%%%%%%%%%%%%%%%%%%%%%%%%%%%%%%%%%%%%%%%%%%%%%%%%%%%%%%%%%
\subsection{Stellar velocity dispersion}\label{sec:stellar_vdisp}
We use the Penalized Pixel-Fitting method (\texttt{pPXF}) developed by \cite{Cappellari2004,Cappellari2017}
to measure stellar velocity dispersion ($\sigma_*$). We used a python wrapper developed in
the course of the PHANGS-MUSE survey by F. Belfiore and I. Pessa \citep{Belfiore2022,Emsellem2022}
and based on the gist package (Galaxy IFU Spectroscopy Tool; \citealt{Bittner2019}).
As required by \texttt{pPXF}, we first resample the MUSE data to a logarithmic wavelength axis
using a channel size of 50\, km\,s$^{-1}$.
Following \cite{Emsellem2022}, this channel size is sufficient to Nyquist sample the
line spread function of MUSE for wavelengths at approximately 7000\,\AA, while
over-sampling it at the blue end.
In order to avoid strong sky residuals the wavelength range for fitting
was limited to 4850--7000\,\AA.
In the following we briefly describe the fitting routine as implemented in
\texttt{pPXF}. For the stellar continuum fitting
the E-MILES simple stellar population models of
\cite{Vazdekis2016} are used in combination with a \cite{Chabrier2003} initial mass function,
BaSTI isochrones \citep{Pietrinferni2004}, eight ages
(0.15--14\,Gyr), and four metallicities ([Z/H] = [-1.5, -0.35, 0.06, 0.4]).
Thus, a total number of 32 templates are used.
Spectral ranges of strong ionised gas emission lines are masked using
a width of $\pm$400\,km\,s$^{-1}$.
Since E-MILES offers a higher resolution than our MUSE data, the
templates are convolved to the spectral resolution of
our data, using an appropriate wavelength-dependent kernel.
We fitted four moments of the line-of-sight velocity distribution:
velocity, velocity dispersion, h3 and h4. To derive the stellar kinematics we make
use of additive Legendre polynomials (12th order, in the spectral direction),
and no multiplicative polynomials. 
The uncertainties on the kinematic parameters are formal errors as given by \texttt{pPXF}.

In the literature, the stellar velocity dispersion in galaxies is often
estimated from the stellar surface density following the prescription of
\cite{Leroy2008}:
\begin{equation}
    \sigma_*\ =\ 1.67\ \sqrt{\frac{2\ \pi\ G\ l_*}{7.3}}\ \Sigma_*^{0.5},
\end{equation}
with $\Sigma_*$ being the observed stellar surface density
in SI units (kg\,m$^{-2}$),
and $l_*$ being the
stellar scale length (=630\,pc measured via fitting of an exponential profile to the data) in $m$.
The underlying assumptions of the equation are the following: the exponential stellar scale height h$_*$
of the galaxy does not vary with radius, and $h_*$ is related to the stellar scale
length $l_*$ via $l_*$/$h_*$=7.3$\pm$2.2, i.e. the flattening ratio measured by \cite{Kregel2002}.
It is further assumed that the disc is isothermal in the z-direction and
hydrostatic equilibrium then allows one to derive $\sigma_*$ from the observed stellar surface density $\Sigma_*$
and the estimated stellar scale height. Finally a fixed ratio
of 0.6
between the radial and vertical component of the
velocity dispersion is assumed, which is reasonable for most late-type galaxies \citep{Shapiro2003}.
We refer the reader to the appendix of \cite{Leroy2008} for more details.

\begin{figure}
	\includegraphics[width=\columnwidth]{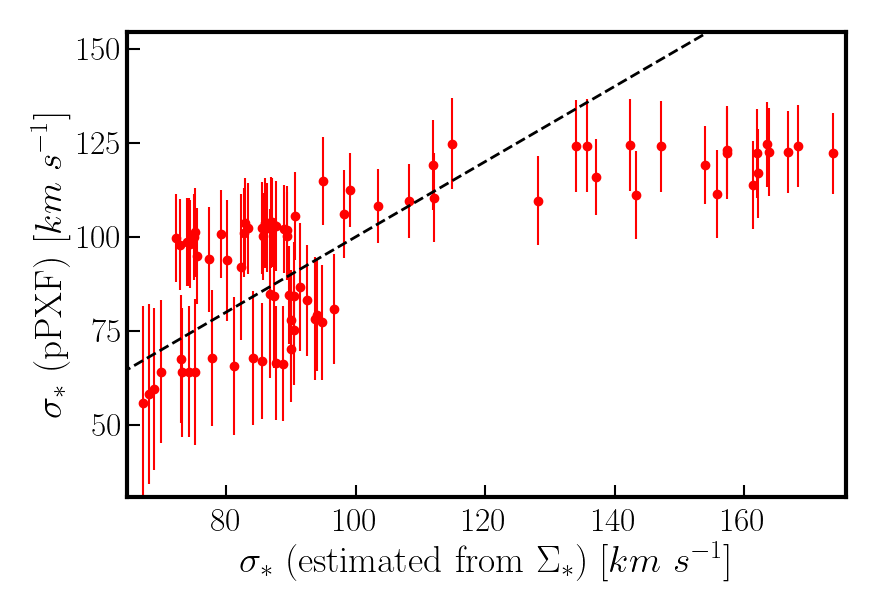}
    \caption{Comparison of radial averages of the velocity dispersion derived from MUSE (\textit{y-axis})
    and estimated using the prescription of Leroy et al. (\textit{x-axis}). The vertical errorbars show the propagated
    uncertainties of the measurements in each radial ring.}
    \label{fig:vdisp_comparison}
\end{figure}

A comparison of radial averages of the velocity dispersion derived from MUSE and estimated as explained above
is shown in Figure \ref{fig:vdisp_comparison}. We find for LARS~8 that in particular in the central region
where the velocity dispersion is highest, the estimates overshoot the true (MUSE-based) values by up to
approximately 40 percent.

%%%%%%%%%%%%%%%%%%%%%%%%%%%%%%%%%%%%%%%%%%%%%%%%%%%%%%%%%%%%%%%%%%%%
%%%%%%%%%%%%%%%%%%%%%%%%%%%%%%%%%%%%%%%%%%%%%%%%%%%%%%%%%%%%%%%%%%%%
\subsection{Toomre $Q$ disc stability}\label{sec:toomreQ}
As gravitational instability is believed to hold a key part in driving gas turbulence \citep{Agertz2009a,Krumholz2016},
we consider a theoretical framework to evaluate this instability.
One of the most common ways of quantifying this instability is Toomre's $Q$ parameter \citep{Toomre1964},
which governs the stability of a smaller patch inside a disc system.
%Toomre's Q parameter was first derived by \cite{Safronov1960}, but got its name after \cite{Toomre1964},
%in which Toomre derived and presented this parameter in a more general and comprehensive way.
The Toomre parameter for an axisymmetric, fluid disc with a differential rotation, can be
determined by analysing the response of the disc to a small perturbation.
The growth of this perturbation
is driven by gravity, expressed as a surface density wave function.
By evaluating the dispersion relation,
first shown by \cite{Safronov1960},
\cite{Toomre1964} found the condition:
\begin{equation}\label{eqn:toomre_gas}
    Q_g\ =\ \frac{\kappa\ \sigma_g}{\pi\ G\ \Sigma_g} > 1
\end{equation}
for the disc being locally stable against graviational collapse.
In this work we compute the epicyclic frequency $\kappa$
as $1.41\ \frac{v(r_\mathrm{gal})}{r_\mathrm{gal}}\ \sqrt{1+\beta}$ and $\beta\ =\ \frac{d \log\ v(r_\mathrm{gal})}{d \log\ (r_\mathrm{gal})}$.
For $Q_g$<1, the disc is locally unstable. In this equation, $\kappa$ is the epicyclic frequency,
$\Sigma_g$ is the gas surface density, $\sigma_g$ is the gas velocity dispersion
(from the rotation curve analysis) and $G$ is the gravitational constant.
The physical meaning of $\kappa$ can be thought of as the rotational support against collapse,
$\sigma$ is the pressure support against collapse and $\Sigma$
sets the level of self-gravity driving the instability.
An implementation of the method to compute $Q_g$ is provided via GitHub by \cite{ToomreCode}\footnote{\url{https://github.com/astrojohannes/toomreQ}},
including a working example.
The above method can be expanded to a disc filled with star particles and differ only slightly from the approach of a fluid:
\begin{equation}\label{eqn:toomre_star}
\centering
    Q_*\ =\ \frac{\kappa\ \sigma_*}{3.36\ G\ \Sigma_*}
\end{equation}

Combining the Toomre parameters for stars and gas is a necessary step to determine the stability of a multi-component disc, 
which is the case for most galaxies. We assume that $\kappa$ is the same for both the gaseous and the stellar disc,
i.e. that gas and stars follow the same rotation.
We stress that the combined $Q$ is derived such that it also obeys the instability
criterion of $Q\sim$1.
There have been several different approaches to combining $Q$ parameters and an extensive look
into different methods was done by \cite{Romeo2013}. 
In this paper, we use the approximation of \cite{Romeo2013}:

\noindent For $Q_* > Q_g$ (gas dominated regime):
\begin{equation}
    \frac{1}{Q} = \frac{1}{Q_g} + \frac{CF}{Q_*}
\end{equation}
For $Q_* <= Q_g$ (star dominated regime):
\begin{equation}
    \frac{1}{Q} = \frac{CF}{Q_g} + \frac{1}{Q_*}
\end{equation}
The correction factor CF is given for both cases via:
\begin{equation}
    \frac{2 \sigma_* \sigma_g}{\sigma^2_* + \sigma^2_g}
\end{equation}

%%%%%%%%%%%%%%%%%%%%%%%%%%%%%%%%%%%%%%%%%%%%%%%%%%%%%%%%%%%%%%%%%%%%
%%%%%%%%%%%%%%%%%%%%%%%%%%%%%%%%%%%%%%%%%%%%%%%%%%%%%%%%%%%%%%%%%%%%
\subsection{Uncertainty of the Toomre Q parameter measurement}
In order to assess the uncertainty of the derived Toomre $Q$
parameter, we propagate our measurement uncertainties, i.e.
the final variance of the Toomre $Q_g$ parameter is
given by the sum of the following products:
\begin{itemize}
    \item squared partial derivatives of $Q_g$ with respect to $\kappa$ times the square of the propagated uncertainty of $\kappa$,
    \item squared partial derivatives of $Q_g$ with respect to $\sigma_g$ times the square of the uncertainty of $\sigma_g$,
    \item squared partial derivatives of $Q_g$ with respect to $\Sigma_g$ times the squared uncertainty of $\Sigma_g$.
\end{itemize}
The uncertainty of $Q_*$ is calculated in an analogue way, but with derivatives of $Q_*$ with respect to $\sigma_*$
and $\Sigma_*$ and using the uncertainties on these parameters.
For reference, we show the exact formulas we used below:
\begin{equation}
    \mathrm{var}(Q_g) = \frac{\sigma_g^2\ \kappa^2\ \mathrm{unc}(\Sigma_g^2)}{\pi^2\ G^2 \Sigma_g^4} +
    \frac{\sigma_g^2\ \mathrm{unc}(\kappa^2)}{\pi^2\ G^2\ \Sigma_g^2} + 
    \frac{\kappa^2\ \mathrm{unc}(\sigma_g^2)}{\pi^2\ G^2\ \Sigma_g^2}
\end{equation}
\begin{equation}
    \mathrm{var}(Q_*) = \frac{\sigma_*^2\ \kappa^2\ \mathrm{unc}(\Sigma_*^2)}{3.36^2\ G^2 \Sigma_*^4} +
    \frac{\sigma_*^2\ \mathrm{unc}(\kappa^2)}{3.36^2\ G^2\ \Sigma_*^2} + 
    \frac{\kappa^2\ \mathrm{unc}(\sigma_*^2)}{3.36^2\ G^2\ \Sigma_*^2}
\end{equation}
\begin{equation}
    \mathrm{var}(\kappa) = \frac{1.9881\ \mathrm{unc}(v(r_{\mathrm{gal}}))^2}{r_{\mathrm{gal}}^2}
\end{equation}
\noindent Uncertainty for $Q_* > Q_g$ (gas dominated regime):
\begin{equation}
    \mathrm{var}(Q_{\mathrm{tot}}) = \frac{CF^2\ \mathrm{unc}(Q_*)^2}{Q_*^4\ \big( \frac{CF}{Q_*} + \frac{1}{Q_{\mathrm{gas}}} \big)^4} + 
    \frac{\mathrm{unc}(Q_g)^2}{Q_{\mathrm{gas}}^4\ \big( \frac{CF}{Q_*} + \frac{1}{Q_{\mathrm{gas}}} \big)^4}
\end{equation}
\noindent Uncertainty for $Q_* <= Q_g$ (star dominated regime):
\begin{equation}
    \mathrm{var}(Q_{\mathrm{tot}}) = \frac{CF^2\ \mathrm{unc}(Q_\mathrm{gas})^2}{Q_\mathrm{gas}^4\ \big( \frac{CF}{Q_\mathrm{gas}} + \frac{1}{Q_*} \big)^4} + 
    \frac{\mathrm{unc}(Q_*)^2}{Q_*^4\ \big( \frac{CF}{\mathrm{gas}} + \frac{1}{Q_*} \big)^4}
\end{equation}

The individual uncertainties that occur in the equations above are
estimated in the following way. Uncertainties of
the kinematics parameters derived with \texttt{$^{3D}$BAROLO}
are obtained via exploration of the parameter space around the
best fit solutions using an MCMC approach \citep{Iorio2017}. Hence, the
uncertainties for the gas velocity dispersion, rotation velocity
and $\kappa$ should be robust
and statistically significant measures.

We stress that asymmetric drift correction is negligible in our case,
because the rotation velocities (see Table \ref{tab:barolo})
of LARS~8 are more than ten times higher than the velocity dispersions
\citep{deBlok2008,Iorio2017}.
The uncertainties on the final rotation curve are thus equal
to the uncertainties of the rotation velocity.

Please refer to the individual sections on gas and stellar masses
for a description of how their uncertainties were estimated.

%%%%%%%%%%%%%%%%%%%%%%%%%%%%%%%%%%%%%%%%%%%%%%%%%%%%%%%%%%%%%%%%%%%%
%%%%%%%%%%%%%%%%%%%%%%%%%%%%%%%%%%%%%%%%%%%%%%%%%%%%%%%%%%%%%%%%%%%%
\subsection{Molecular clumps: identification, virial mass and virial parameter}\label{sec:clump_identification}
%see also Colombo Thesis:
%https://www.imprs-hd.mpg.de/54475/thesis_colombo.pdf
We apply \texttt{CPROPSTOO} \citep{Williams1994,CPROPSCODE,Leroy2015}, an IDL package that is available
through GitHub\footnote{\url{https://github.com/akleroy/cpropstoo}} and was
developed to identify and measure properties of molecular clouds or clumps in fits data cubes.
In particular, \texttt{CPROPSTOO} corrects for the effects of beam convolution and sensitivity when
measuring physical properties such as masses or sizes of identified clouds,
allowing to make unbiased (beam-independent) measurements.
Using a growth-curve analysis on the observed emission line, the
algorithm thus extrapolates the measurements to values one would expect in the
case of perfect sensitivity. Additionally, \texttt{CPROPSTOO} corrects for finite
resolution in both the velocity and spatial domain. This is done via
de-convolution of the telescope beam and the width of a spectral channel
from the measured cloud size and line width.
For more details, we refer the reader to the aforementioned publications.
Here we report the main parameters for the \texttt{find\_local\_max} task
that we applied for clump identification:
delta=2, /snr, minpix=20, minarea=2, minvchan=2, friends=4, specfriends=2.

% rephrase
The virial mass depends on measurements of the size and the observed
line width (due to turbulence) of the cloud. If these quantities are known and the radial density profile is given,
then following \cite{Solomon1987}, the mass of the cloud under the assumptions of virial equilibrium and
spherical symmetry can be calculated:
\begin{equation}
    M_\mathrm{vir} = \frac{3(5-2\gamma)}{G(3-\gamma)} \Delta v^2 R
\end{equation}

The virial mass $M_\mathrm{vir}$ is then given in M$_{\odot}$ and depends on the radial density distribution exponent $\gamma$,
the linear cloud size (R) in parsec and the full width at half-maximum (FWHM) of the line in km/s ($\Delta$v).
Taking the frequently assumed $\gamma$=1 radial
density distribution exponent, corresponding to a cloud radial density profile $\rho\ \propto\ r^{-1}$ \citep{Maclaren1988,Hughes2010},
the above equation can be re-written as:
\begin{equation}\label{eqn:Mvir}
    M_\mathrm{vir} = 1040 \sigma^2 R
\end{equation}
The units are M$_{\odot}$, km\,s$^{-1}$ and pc for M$_\mathrm{vir}$, $\sigma$ and $R$ respectively.
In this equation the numerical coefficient accounts for the radial density profile,
the conversion factor between FWHM and velocity dispersion
($\Delta$v=2.35$\sigma$) and the gravitational constant.
% from Jiayi 2018
Departures from virial equilibrium can be expressed via the virial parameter,
$\alpha_\mathrm{vir}$:
\begin{equation}\label{eqn:alpha_vir}
    \alpha_\mathrm{vir} = \frac{2K}{U} = \frac{5 \sigma^2 R}{G M_\mathrm{lum}} = 1.12 \frac{M_\mathrm{vir}}{M_\mathrm{lum}}
\end{equation}
where $K$ and $U$ denote the kinetic energy and self-gravitational potential energy respectively.
The quantity M$_\mathrm{lum}$ is the luminous molecular mass converted from
low-J CO intensities using a conversion factor.
Virialized clouds without surface pressure or magnetic support have $\alpha_\mathrm{vir}$=1, while
both marginally bound clouds and clouds in free-fall collapse share
energy equipartition ($K=U$) and thus have $\alpha_\mathrm{vir}\sim$2
\citep{Ballesteros-Paredes2011,Camacho2016,Ibanez-Mejia2016,Sun2018}.

%%%%%%%%%%%%%%%%%%%%%%%%%%%%%%%%%%%%%%%%%%%%%%%%%%%%%%%%%%%%%%%%%%%%
%%%%%%%%%%%%%%%%%%%%%%%%%%%%%%%%%%%%%%%%%%%%%%%%%%%%%%%%%%%%%%%%%%%%
\subsection{Dynamical equilibrium pressure}\label{sec:pde}
% See Molly's paper:
% https://arxiv.org/pdf/1803.10785.pdf
%
% Implementation:
%stars_scalelength=630.0   # pc
%stars_scaleheight=stars_scalelength/7.3     # following Kregel et al. (2002) and Leroy et al 2008
%
%# conversion to SI
%
%# lengths to meter
%stars_scalelength*=u.parsec.to(u.m)
%stars_scaleheight*=u.parsec.to(u.m)
%
%# surface densities to kg/m2
%mh2*=u.solMass*u.parsec**(-2)
%mh2=mh2.to(u.kg*u.m**(-2))
%
%stars*=u.solMass*u.parsec**(-2)
%stars=stars.to(u.kg*u.m**(-2))
%
%# linewidths to m/s
%wco*=u.km*u.s**(-1)
%wco=wco.to(u.m*u.s**(-1))
%
%rho_star=stars/(2*stars_scaleheight)    #  following van der Kruit (1988)
%
%# calculate PDE (see equation 7 in Gallagher et al 2018)
%t1= (np.pi*const.G*mh2**2)/2
%t2= mh2 * (2*const.G*rho_star)**0.5 * wco
%pde=t1.value+t2.value
%# convert from kg/m2 --> K cm^-3 using relation: p/k=n*T
%pde /= const.k_B.value    # K m^-3
%pde *= 1e-6               # K cm^-3
%
ISM pressure plays a crucial role in many theories of star formation (e.g. \citealt{Ostriker2011}),
as it determines the gas density distribution \citep{Helfer1997,Usero2015,Bigiel2016,Gallagher2018}.
%In high pressure regions the
%mean star-forming gas density is shifted towards higher values, compared
%to regions of lower dynamical equilibrium pressure.
Following \cite{Elmegreen1989} we estimate the mid-plane dynamic equilibrium
pressure, $P_{\mathrm{de}}$, using the following prescription:
\begin{equation}
    P_{\mathrm{de}}\ =\ \frac{\pi\ G\ \Sigma_{\mathrm{gas}}^2}{2}\ +\ \Sigma_{\mathrm{gas}}\ \sqrt{2\ G\ \rho_*}\ \sigma_{\mathrm{gas}}
\end{equation}
Here, $\Sigma_{\mathrm{gas}}$ is the total gas surface density, including the atomic
and molecular component. Since our study only covers the central part of the
galaxy, i.e. the high-density regime, in which most atomic gas is readily
converted to molecular gas, we may only consider $\Sigma_{\mathrm{H_2}}$ instead.
The vertical velocity dispersion of the gas is denoted as $\sigma_\mathrm{gas}$ and
the parameter $\rho_*$ is the mass volume density of stars and dark matter
at the mid-plane, which we estimate following \cite{vanderKruit1988} using the relation: $\rho_*=\Sigma_*/(2\ h_*)$,
with the disc scale height $h_*$.
$P_{\mathrm{de}}$ then expresses the pressure needed to balance the vertical gravity on the 
gas in the galaxy disc. The first term reflects the gas self-gravity, the second
term reflects the weight of the gas in the potential well of the stars.
Since the stellar potential in LARS~8 exceeds the gas self-gravity,
we expect the second term to be dominant.

%%%%%%%%%%%%%%%%%%%%%%%%%%%%%%%%%%%%%%%%%%%%%%%%%%%%%%%%%%%%%%%%%%%%
%%%%%%%%%%%%%%%%%%%%%%%%%%%%%%%%%%%%%%%%%%%%%%%%%%%%%%%%%%%%%%%%%%%%
%%%%%%%%%%%%%%%%%%%%%%%%%%%%%%%%%%%%%%%%%%%%%%%%%%%%%%%%%%%%%%%%%%%%
%%%%%%%%%%%%%%%%%%%%%%%%%%%%%%%%%%%%%%%%%%%%%%%%%%%%%%%%%%%%%%%%%%%%
\section{Results}\label{sec:results}
The methods outlined in the previous section enable us to
quantify clump/cloud properties (Sections \ref{sec:clumps}--\ref{sec:mass_size_relation})
as well as star formation relations and radial trends in LARS~8 (Section \ref{sec:radial_profiles}).
Finally, the gravitational instability of the disc is shown in Section
\ref{sec:disc_stability}.

%%%%%%%%%%%%%%%%%%%%%%%%%%%%%%%%%%%%%%%%%%%%%%%%%%%%%%%%%%%%%%%%%%%%
%%%%%%%%%%%%%%%%%%%%%%%%%%%%%%%%%%%%%%%%%%%%%%%%%%%%%%%%%%%%%%%%%%%%
\subsection{Identification of molecular clumps in LARS 8}\label{sec:clumps}
\begin{figure*}
	\includegraphics[width=\textwidth]{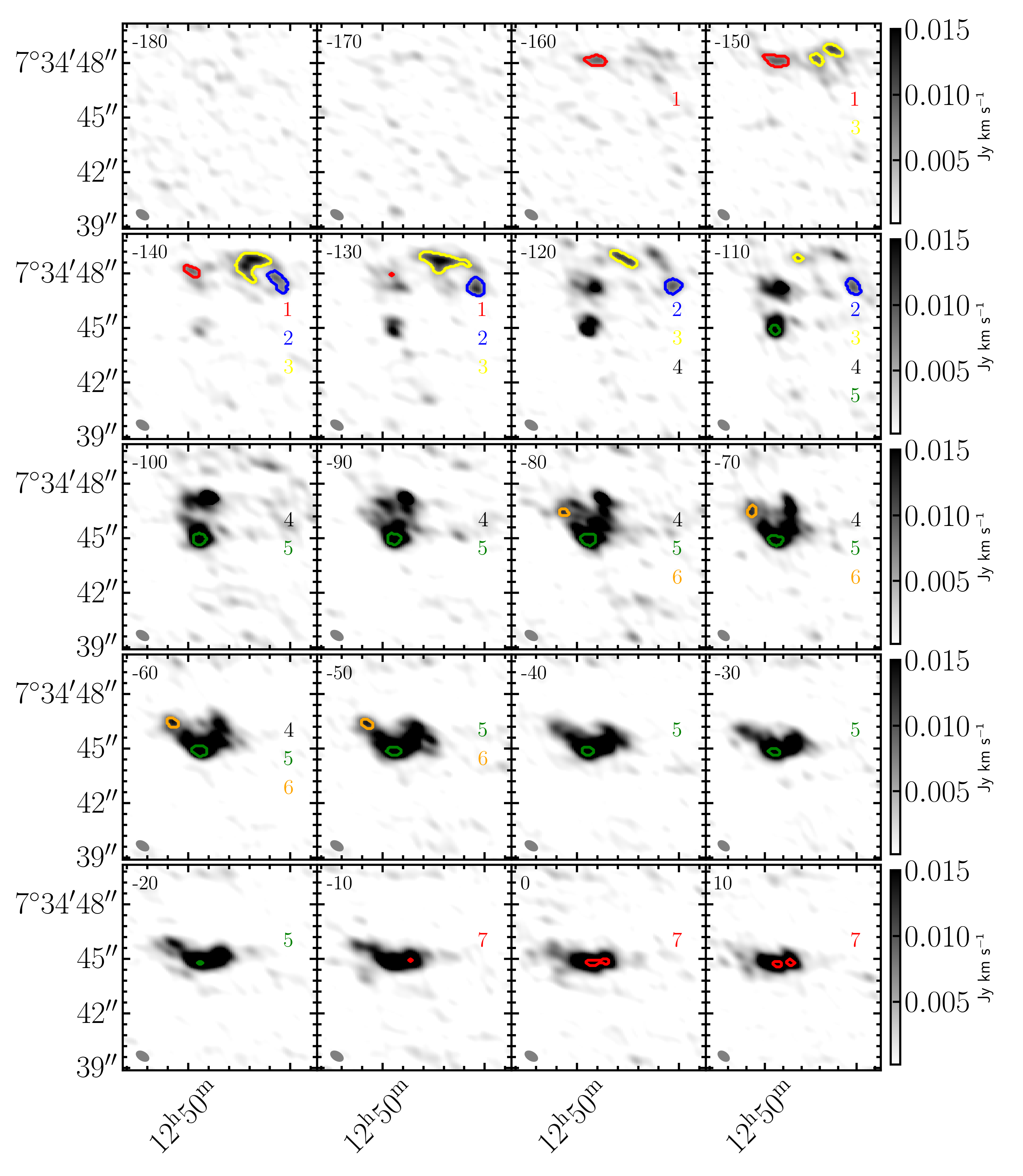}
    \caption{NOEMA CO (2--1) channel maps of the northern part of LARS~8, showing velocities between -180 and +10 km/s with identified \texttt{CPROPSTOO} structures
    (shown as contours). Properties of the identified clumps are summarized in Table \ref{tab:clumpprops}.}
    \label{fig:noema_channels1}
\end{figure*}

\begin{figure*}
	\includegraphics[width=\textwidth]{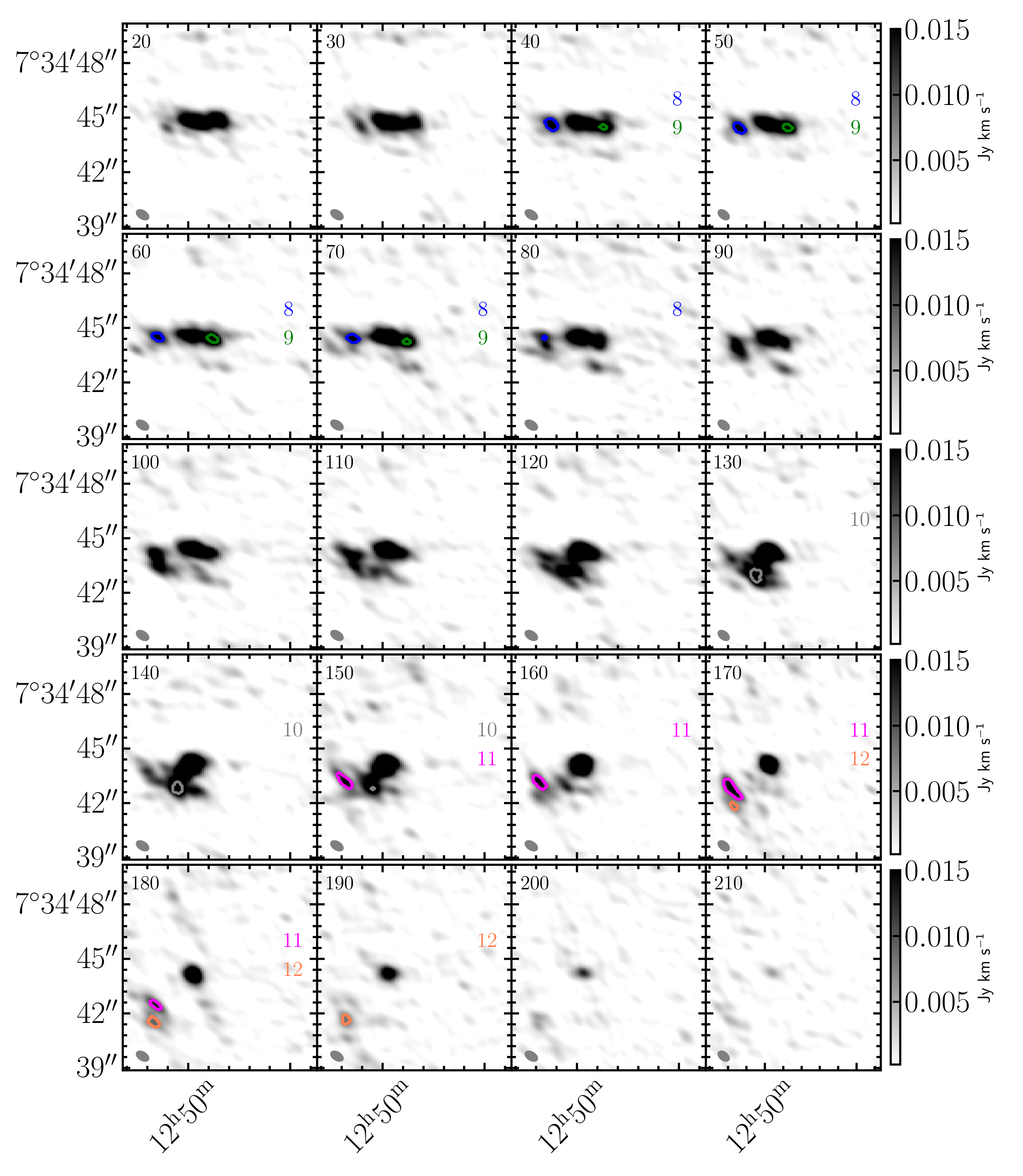}
    \caption{NOEMA CO (2--1) channel maps of the southern part of LARS~8, showing velocities between 20 and +210 km/s with identified \texttt{CPROPSTOO} structures
    (shown as contours). Properties of the identified clumps are summarized in Table \ref{tab:clumpprops}.}
    \label{fig:noema_channels2}
\end{figure*}

Applying \texttt{CPROPSTOO} on our native resolution NOEMA CO (2--1) data cube with a
channel width of 10~km/s, we could identify 12 molecular clumps in total
(see Figures \ref{fig:noema_channels1} and \ref{fig:noema_channels2}).
The unbiased properties of the identified molecular clumps are summarized in Table
\ref{tab:clumpprops}. Their masses
range from 10$^{8.1}$ to 10$^{9.3}$~M$_\odot$, covering
linear (extrapolated) diameters between $\sim$600--2000\,pc. Clump 7 was found
to be the most massive one, located in the very center of the galaxy.
The channel maps between approximately $-$180 km\,s$^{-1}$ and $+$210\,km\,s$^{-1}$ in Figure \ref{fig:noema_channels2}
may suggest that \texttt{CPROPSTOO} failed in associating extended gas to clumps in the central region of the galaxy.
This is because the linewidth in the central few pixels of the galaxy is extremely wide (more than 300\,km$^{-1}$) due to beam smearing.
Much higher resolution (spatially and spectrally) would be needed to identify individual clumps in that part of the galaxy.

\begin{table}
\caption{Properties of the identified molecular clumps. The clumps in Figures \ref{fig:noema_channels1} and
\ref{fig:noema_channels2} are identified by the IDs as given in the Table below.
See also Equations \ref{eqn:Mvir} and \ref{eqn:alpha_vir}.
}\label{tab:clumpprops}
\begin{tabular}{rrrrrr}
\hline \hline
ID & $R_\mathrm{eff}$  & $\log\ M_\mathrm{lum}$  & $\alpha_\mathrm{vir}$ & $\sigma$ & $v_\mathrm{pos}$ \\
   & [pc]       & M$_{\odot}$    &            & [km s$^{-1}$] & [km s$^{-1}$] \\
1  & 601      & 8.20     & 1.2        & 16.6       & -151       \\
2  & 322      & 8.14     & 1.0        & 19.2       & -126       \\
3  & 913      & 8.55     & 0.9        & 16.8       & -135       \\
4  & 343      & 8.59     & 1.0        & 30.7       & -92        \\
5  & 340      & 9.23     & 0.8        & 59.7       & -73        \\
6  & 320      & 8.40     & 1.4        & 31.1       & -63        \\
7  & 1023     & 9.26     & 0.5        & 27.8       & 4          \\
8  & 469      & 8.51     & 1.7        & 32.0       & 53         \\
9  & 436      & 8.76     & 1.1        & 34.5       & 56         \\
10 & 609      & 8.89     & 0.4        & 20.5       & 135        \\
11 & 480      & 8.57     & 1.0        & 26         & 164        \\
12 & 436      & 8.21     & 1.9        & 24.7       & 181        \\
\end{tabular}
\end{table}

%%%%%%%%%%%%%%%%%%%%%%%%%%%%%%%%%%%%%%%%%%%%%%%%%%%%%%%%%%%%%%%%%%%%
%%%%%%%%%%%%%%%%%%%%%%%%%%%%%%%%%%%%%%%%%%%%%%%%%%%%%%%%%%%%%%%%%%%%
\subsection{Mass-size relation for the massive clumps}\label{sec:mass_size_relation}
\begin{figure}
	\includegraphics[width=\columnwidth]{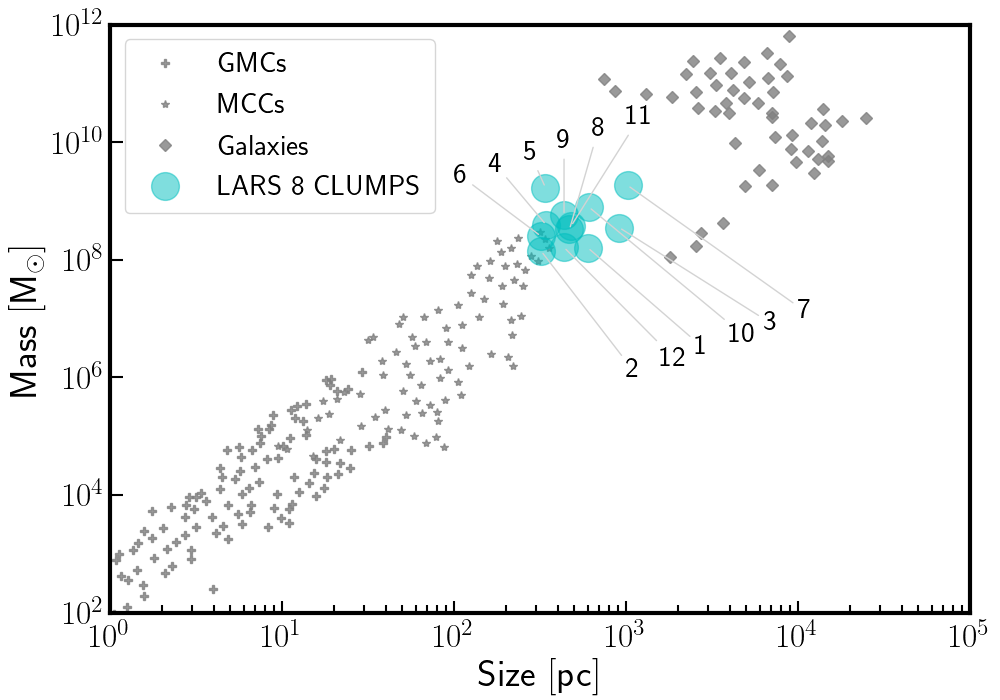}
    \caption{Mass-size relation for the massive clumps identified in LARS~8 (\textit{cyan points})
    with the ID numbers as given in Table \ref{tab:clumpprops}. The clumps are compared to the
    literature compilation of
    \protect\cite{NguyenLuo2016},
    that is based
    on GMC (\textit{plus signs}) data of
    \protect\cite{Onishi2002,Heyer2009,Maruta2010,RomanDuval2010,Evans2014,Shimajiri2015},
    MCC (\textit{stars}) data of
    \protect\cite{Rosolowsky2007,Murray2011,Wei2012,Miura2012,Miura2014,DonovanMeyer2013,Garcia2014}
    and galaxies (\textit{diamonds}) from \protect\cite{Leroy2013,Tacconi2013,Genzel2010}.
    }
    \label{fig:mass_size_rel}
\end{figure}

% See Cava+ (2017)
% https://arxiv.org/pdf/1711.03977.pdf
Figure \ref{fig:mass_size_rel} compares the derived masses and sizes of the
molecular clumps identified in LARS~8 to the literature compilation of \cite{NguyenLuo2016}, which contains
giant molecular clouds (GMCs) of the Milky Way with sizes smaller than 10~pc,
molecular cloud complexes (MCCs) with sizes between 10 and 1000~pc, as well as galaxies
and structures larger than 1~kpc typically found at high redshift.
Note that the identified structures or clumps in LARS~8 are resolved, i.e.
their deconvolved diameters are at least as wide as the beam major axis.
We thus conclude from Figure \ref{fig:mass_size_rel} that
the clumps of diffuse molecular gas in LARS~8 are in fact scaled-up versions
of the MCCs in the literature.
In the mass-size relation they
populate the range between MCCs and structures identified at high redshifts.
However -- despite an ongoing massive star formation process in LARS 8 -- the clumps
follow the same trend between mass and size.
This finding implies a universal (constant) diffuse molecular mass surface density, even in highly star-forming galaxies
such as LARS~8. The elevated star formation rates must thus result from
processes \textit{within} the large diffuse molecular
reservoirs we identified in CO~(2--1). It might be that
either the structures contain more over-densities (e.g. traced by HCN)
or that the star formation is in some way more efficient.
The latter is supported by observations of \cite{Messa2019}, who
have derived sizes and properties of
clumps identified from very high-resolution UV photometry.
They find that the range of clump sizes
in LARS~8 is similar to those in normal star forming galaxies
or at high redshift, i.e. 15--200\,pc.
However, the star formation rates per UV clump are higher
and fall between those observed in the local and high-$z$ Universe.
Also, a combination of both more dense clumps and higher efficiency per clump may apply.

%%%%%%%%%%%%%%%%%%%%%%%%%%%%%%%%%%%%%%%%%%%%%%%%%%%%%%%%%%%%%%%%%%%%
%%%%%%%%%%%%%%%%%%%%%%%%%%%%%%%%%%%%%%%%%%%%%%%%%%%%%%%%%%%%%%%%%%%%
\subsection{Radial profiles and KS relation}\label{sec:radial_profiles}
\begin{figure}
	\includegraphics[width=\columnwidth]{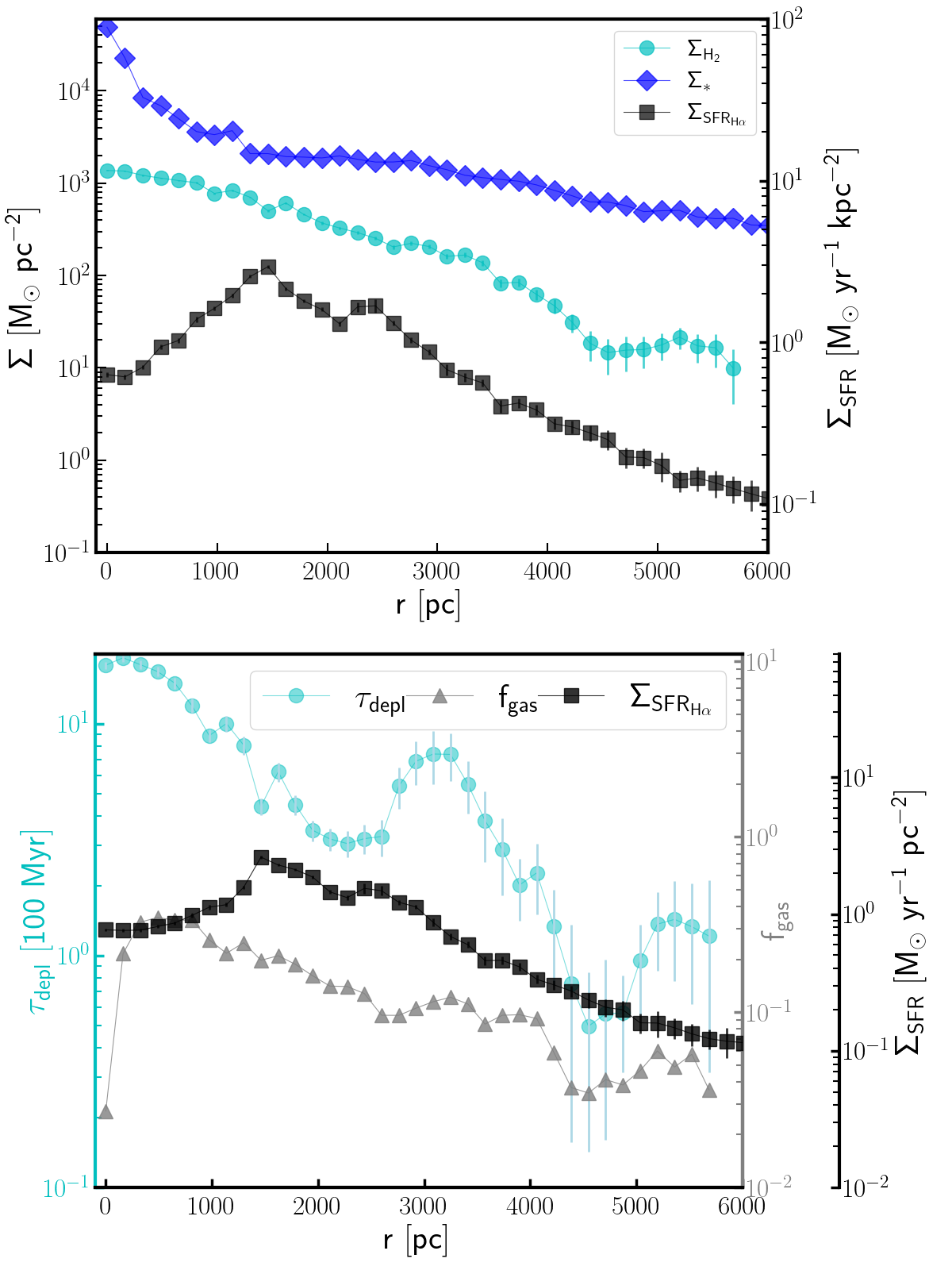}
    \caption{\textit{Top panel}: Elliptical inclination-corrected profiles of stellar (\textit{blue diamonds}), molecular (\textit{cyan circles})
    and SFR surface densities (\textit{black squares}).
    \textit{Bottom panel}: Same as top panel for the molecular gas depletion time (\tdep), the H$\upalpha$-based
    star formation rate surface density ($\Sigma_\mathrm{SFR}$) and the molecular gas fraction ($f_\mathrm{gas}$).}
    \label{fig:radprof}
\end{figure}

Figure \ref{fig:radprof} shows inclination-corrected, elliptical profiles 
of several quantities we have derived, centered on the maximum stellar surface density.
The plots show that while the molecular gas surface
density declines relatively smoothly from the center outwards, the
stellar surface density is peaked in the innermost $\sim$500\,pc. This
peak may represent a bulgelike structure that is about to form, similar to
observations in high redshift discs \citep{Elmegreen2009} and
as predicted by numerical simulations, e.g. in \citet{Elmegreen2008}.
Thereby, gas-rich disc galaxies show disc instabilities that first trigger clump formation.
These clumps (and other disc matter) move inwards and merge, forming a bulge
(or bulgelike-clump) that is characterized by a
Sersic index n=4 (like a classical bulge) and rotation.
See \cite{Rasekh2022} for a compilation of Sersic profiles for LARS galaxies.

In contrast, the star formation rate density
is highest in a ring-like structure located at a radius of $\sim$1.2~kpc.
The lowering of the SFR towards the inner kiloparsec in combination with the
low molecular gas fraction suggests that some process has quenched star formation
in the center, e.g. AGN feedback.
Alternatively, it might be that the extinction correction underestimates the
true SFRs in the innermost parts, where H$\upalpha$ becomes optically thick. However,
this would not explain the relatively low gas fraction in the center.

Moreover, Figure \ref{fig:radprof} reveals that the molecular gas depletion
time, \tdep, strongly declines from more than 1~Gyr in the center to $\sim$100\,Myr in
the outer parts of the disc. This contrasts normal star-forming galaxies that
typically have roughly constant \citep{Bigiel2011} or even radially increasing gas depletion time
scales \citep{Leroy2008}.
This behaviour is further suggested by (some) gravity-driven theoretical models of star formation,
e.g. \cite{Krumholz2012} argue that in the regime of normal star-formation the
GMCs are basically decoupled from the rest of the ISM. The depletion
time is then mainly set by the internal properties and processes of the GMCs -- that are roughly
constant in normal Milky-Way-like clouds -- rather than by the large-scale behavior of the ISM.
\cite{Krumholz2012} further argue that in starbursts (with a Toomre~$Q$ parameter $\sim$1)
the depletion time should be set by the orbital (dynamical) time.
However, given the fact that the orbital time increases with radius (flat rotation) one would expect
from such theory that the depletion time increases with radius. This is not observed
in LARS~8.

\begin{figure}
	\includegraphics[width=\columnwidth]{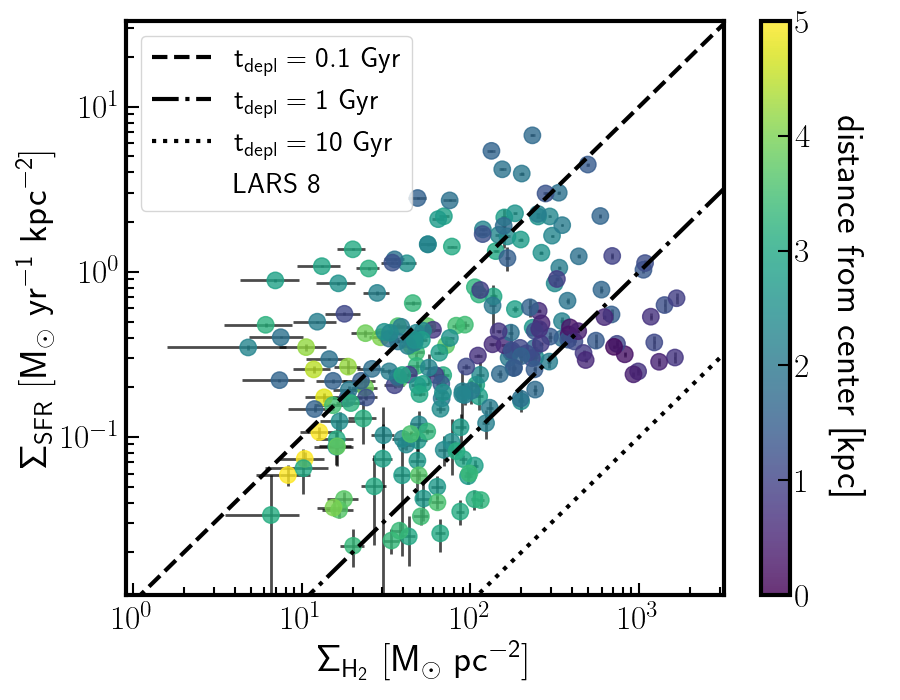}
    \caption{Resolved molecular Kennicutt-Schmidt relation for LARS~8, based on SFRs from extinction-corrected H$\upalpha$ and molecular
    masses from CO~(2--1), both corrected for inclincation. Each point corresponds to a measurement in a $\sim$650\,pc sized region/pixel.
    The colors indicate distance from the center.}
    \label{fig:ks}
\end{figure}

The molecular Kennicutt-Schmidt relation for LARS 8 is presented in
Figure \ref{fig:ks}.
Each point in the plot represents an independent measurement (line-of-sight) that
we calculated from the mean value within 2x2 bins (using \texttt{numpy} \texttt{reshape}).
It is seen that the measurements of
individual lines-of-sight exhibit a relatively large scatter
within a range of roughly one order of magnitude.
However, the central region forms an interesting feature that is
characterised by a roughly constant star formation rate density
while the molecular gas surface density varies by up to an order of magnitude,
with a mean gas depletion time around $\sim$1\,Gyr.

%\begin{figure}
	%\includegraphics[width=\columnwidth]{plots/ks_tau_ff.png}
    %\caption{Resolved Kennicutt-Schmidt relation for LARS 8, normalize to the free-fall time.}
    %\label{fig:ks_tff}
%\end{figure}

%%%%%%%%%%%%%%%%%%%%%%%%%%%%%%%%%%%%%%%%%%%%%%%%%%%%%%%%%%%%%%%%%%%%
%%%%%%%%%%%%%%%%%%%%%%%%%%%%%%%%%%%%%%%%%%%%%%%%%%%%%%%%%%%%%%%%%%%%
%\subsection{Spatial Distribution of Ionized and Molecular Gas at $\sim$400pc scale}
% Krujssen, Chevance: Nature (CO/Ha) offset
% Deanne Fisher: turbulent gas discs --> decrease/gradient in SFE with radius
% Note: observations of CO at large radii difficult --> uv abs lines

%%%%%%%%%%%%%%%%%%%%%%%%%%%%%%%%%%%%%%%%%%%%%%%%%%%%%%%%%%%%%%%%%%%%
%%%%%%%%%%%%%%%%%%%%%%%%%%%%%%%%%%%%%%%%%%%%%%%%%%%%%%%%%%%%%%%%%%%%
\subsection{Disc stability - Toomre $Q$ analysis}\label{sec:disc_stability}
\begin{figure}
	\includegraphics[width=1\columnwidth]{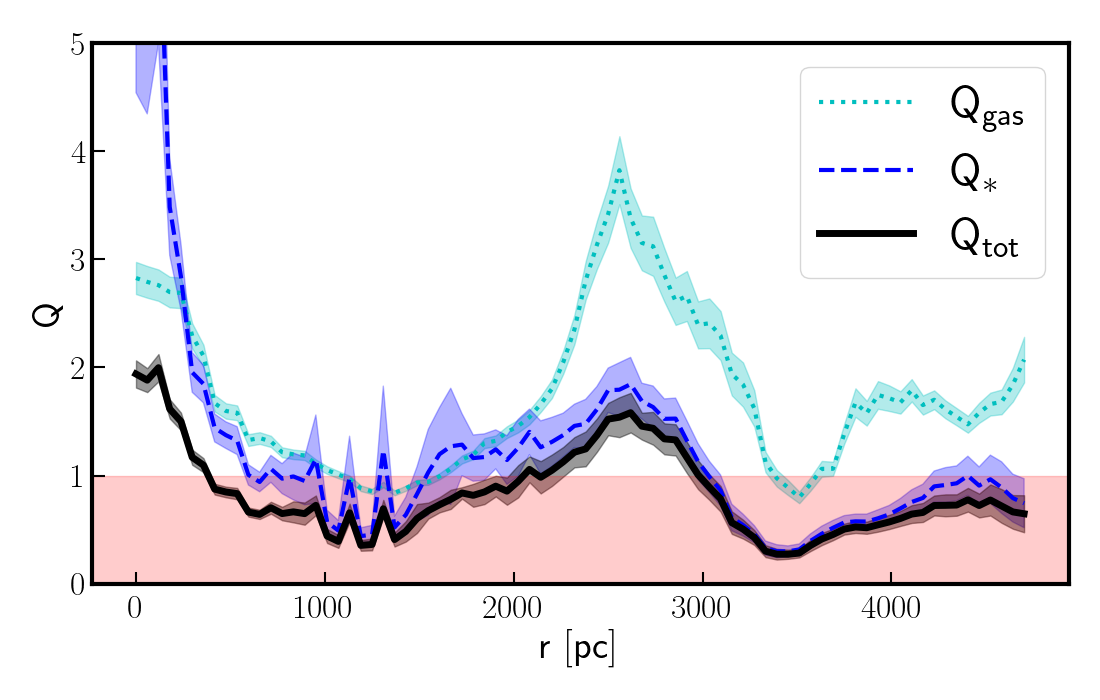}
    \caption{LARS 8 disc stability from radial Toomre $Q$ analysis. Regions with $Q<$1 (\textit{light red shaded area}) are considered unstable.
    The total instability parameter ($Q_\mathrm{tot}$) is shown as \textit{black} curve with the \textit{grey shaded area} indicating its uncertainty.
    The contribution of the stellar and the gaseous component are shown by the \textit{blue} and \textit{cyan} curves respectively.}
    \label{fig:toomreQ}
\end{figure}

Using the smoothed rotation curve (see Figure \ref{fig:rotcurve}) derived from the NOEMA CO~(2--1) data cube,
and subsequent calculation of the $\beta$-parameter and the epicyclic frequency $\kappa$, the Toomre~$Q$
parameters for the molecular gas ($Q_{\mathrm{gas}}$), the  stellar component ($Q_*$) and the combined total
instability parameter ($Q_{\mathrm{tot}}$) could be computed as a function of galactocentric radius (see Figure \ref{fig:toomreQ}).
Note that we have centered the previously discussed radial profiles
on the stellar peak, while here we (have to) use the kinematic center.
Between these two we find an offset of $\sim$0.8\,arcsec or $\sim$650\,pc.
Such offset is also found in numerical simulations of \cite{Elmegreen2008}
during the phase of the formation of a central bulgelike-clump.
We stress that the overall shape of the radial profiles does not
change if the kinematic center is used instead.

Figure \ref{fig:toomreQ} reveals that only the innermost $\sim$500\,pc of LARS~8 are stable.
This central stability is mainly driven by high values of $\kappa$ due to the extremely steep
rise in rotation velocity that causes very high $\beta$ values.
Note that although we cannot
(kinematically) resolve the central $\sim$500\,pc, i.e. we cannot distinguish between rotation
and dispersion (beam smearing), it is still possible to compute $\kappa$.
The plot further shows that the outskirts of the disc are unstable over large scales, with values
of $Q_{\mathrm{tot}}$ well below the critical limit of 1.
Such highly unstable discs are
not observed in normal star-forming disc galaxies \citep{Leroy2008},
but seem typical for the clumps observed in massive high-$z$ discs
\citep{Genzel2011,Wisnioski2012,Mieda2016}.
The relatively high star formation rate surface densities
observed in LARS~8 over large scales are thus likely the result of
enhanced disc fragmentation due to $Q_{\mathrm{tot}}<<$1.
These instabilities thus trigger the formation of massive stellar and molecular clumps.

However, it seems that purely gravity-driven theoretical models of star formation
do not reproduce our observations, in particular e.g. \cite{Krumholz2012} predict for galaxies
in the Toomre regime (as LARS~8) a positive correlation between
the molecular gas depletion time and the orbital period. As explained,
this is not observed in LARS~8.

Other models assume that the star formation process is self-regulated and
thus leads to pressure balance in the ISM. In particular, the star-forming system
is then in balance between feedback processes from star formation and the external pressure.
In case of a disc galaxy the relevant pressure is then $P_{\mathrm{de}}$, the dynamical equilibrium
pressure. Based on that, e.g. \cite{Ostriker2011} predict a linear relation
between the star formation rate surface density and the ISM pressure. We test
this prediction in the next section. 

%%%%%%%%%%%%%%%%%%%%%%%%%%%%%%%%%%%%%%%%%%%%%%%%%%%%%%%%%%%%%%%%%%%%
%%%%%%%%%%%%%%%%%%%%%%%%%%%%%%%%%%%%%%%%%%%%%%%%%%%%%%%%%%%%%%%%%%%%
%%%%%%%%%%%%%%%%%%%%%%%%%%%%%%%%%%%%%%%%%%%%%%%%%%%%%%%%%%%%%%%%%%%%
%%%%%%%%%%%%%%%%%%%%%%%%%%%%%%%%%%%%%%%%%%%%%%%%%%%%%%%%%%%%%%%%%%%%
\section{Discussion}\label{sec:discussion}
% Center-Disc Toomre Q
We showed in the previous section that the galactic disc of LARS~8
is highly unstable, in particular at radii outwards of $\sim$500\,pc.
The Toomre~$Q$ parameter is found to be significantly lower than one (see Figure \ref{fig:toomreQ})
and we conclude that the formation of the observed massive
molecular and stellar clumps is driven by fragmentation of the disc
rather than accretion of external mass or merging. Contrarily,
the central region of LARS~8 was found to be different. It has a Toomre~$Q$ parameter greater than one and is thus stable, it has a relatively low gas fraction and a low star formation rate density,
and it has a depletion time of more than $\sim$1\,Gyr (which is much longer than within the disc).
%The central region might thus be under strong influence of feedback from
%star formation.

% Center --> Shear suppresses  SFR
\cite{Utomo2017} studied the molecular gas depletion time as a function of local
environment in 52 non-AGN disc galaxies drawn from the EDGE-CALIFA \citep{Sanchez2012,Bolatto2017}
survey. They find that galaxies with increased central stellar surface densities
(relative to the disc) typically show a decrease in \tdep\ in the center.
As stellar surface density is the determining factor for ISM pressure, \cite{Utomo2017}
claim that the observed shorter central gas depletion times are a consequence of
higher external pressure that facilitates cloud collapse.
In the center of LARS~8 we also observe an increase in stellar surface density compared
to the disc, but at the same time -- for the center -- we find \textit{longer} molecular gas depletion times.
Additionally, our
radial plots (Figure \ref{fig:radprof})
show that
the star formation rate surface density sharply drops towards the center,
while the molecular gas surface density in LARS~8 decreases
relatively smoothly from the center to the outskirts.

Some process in the center must therefore lead to quenching of star formation.
We find evidence that shear is mainly responsible for the suppression of
star formation in the center, as we see that $\kappa$\ increases by a factor of $\sim$4
between a radius of $\sim$1\,kpc towards the innermost central region.
Also the gas velocity dispersion increases in that radial regime (from $\sim$1\,kpc to the center),
however only by roughly 50 percent. Thus, it is mainly shear that causes $Q>>1$ in the center.
Feedback from supernovae (that drive the gas velocity dispersion) thus plays only a minor role (if any) for
the suppression of star formation.
In fact our kinematic
results (see Table \ref{tab:barolo}) even suggest that the gas velocity dispersion
drops towards the innermost region.
This observation further rules out
feedback from SNe, but we caution that the measurement of velocity dispersion
in the center is relatively uncertain due to beam smearing caused by the steep increase
of the rotation curve. However, further support against star formation quenching due to SNe 
is found from stellar population synthesis performed by Melinder et al (in preparation).
They show that the central $\sim$500\,pc of LARS~8 are dominated by old stars with ages $>$1\,Gyr.

We also compare our observations to the feedback models of
e.g. \cite{Ostriker2010} or \cite{Faucher-Giguere2013}, which are based on
a balance between energy injected through feedback and disc vertical pressure.
The models predict an inverse relation between \tdep\ and
the vertical gas velocity dispersion. Such relation was previously observed by
\cite{Fisher2019} in a set of massive and highly turbulent discs. However, from
our data of the central region of LARS~8 we cannot test any
such correlation, because of spatial resolution and beam smearing 
that makes measurements of the velocity dispersion extremely challenging.

Given our current data,
we thus conclude that shear is the most likely cause
for the relatively low star formation rates in the center and the long depletion times.
As we mentioned in the previous section, it might also be the case that the
computed star formation rates in the center are somewhat spurious due to the relatively
high extinction found in this region.
We stress that the total galaxy-wide SFR from extinction-corrected H$\upalpha$ is in fact 50
percent higher than the SFR we previously derived in \cite{Puschnig2020} from far infrared
measurements. This might be an indication that the Balmer-decrement method
overestimates the true fluxes/SFRs rather than underestimating it. On the other hand,
the discrepancy between infrared and H$\upalpha$ based SFRs may be the result of
a star formation history with a recent burst (to which H$\upalpha$ is more sensitive).

%However, the low gas fraction, long gas depletion time and relatively low
%star formation rate surface density in the center of LARS~8 provides
%evidence that feedback from star formation is the dominant physical process
%that counteracts the high external pressure and self-regulates star formation.
%
% Disc: radial decrease of mol. gas depletion time
In contrast, the environmental properties of the outer disc of LARS~8
are different, in particular we find that the disc is highly unstable.
The radial profiles in Figure \ref{fig:radprof} further revealed that the molecular gas depletion time in LARS~8
decreases with the galactic radius. This behaviour is contrary
to what is typically observed in nearby disc galaxies
in which \tdep\ either stays flat or slightly increases with radius \citep{Leroy2008}.
Note that models of star formation in stable discs, e.g. \cite{Krumholz2012}, predict exactly such behaviour
for GMCs that are basically decoupled from the large-scale ISM.
In these models star formation is mainly dictated by local properties
rather than large-scale effects.
Additionally, \cite{Semenov2017} showed that for regular, local spiral galaxies, \tdep\ is $\sim$1--2\,Gyr
due to the long time the gas spends in the non-star-forming phase, while only a small fraction of the
gas is converted into stars within a short time.

The difference in the radial \tdep\ profiles
between LARS~8 and
normal star forming disc galaxies \citep{Romeo2017}
is thus the result
of the observed large-scale Toomre instabilities in LARS~8,
in which
the ISM is dominated by dozens of supermassive star-forming clouds that
disallow the star forming regions to decouple from the
ambient ISM (as they make up the ISM).
\cite{Krumholz2012} also made predictions of \tdep\
for starbursts in the Toomre regime ($Q\sim$1), for which they
find that \tdep\ should mainly be dictated by the dynamic timescale,
i.e. $2 r \pi/v_\mathrm{rot}$. Our observations, however,
are not in agreement with this prediction of a radially increasing
gas depletion time. We argue that the observed instabilities in LARS~8
are more violent ($Q<<$1) and thus involve more complex physical processes
such as galaxy-scale shocks or inflows \citep{Barnes2004,Teyssier2010,Powell2011}
which were omitted by the models of \cite{Krumholz2012}.
%The preceeding text is very interesting and could be slightly expanded. 
%Semenov et al. 2017 showed that for regular/local spiral galaxies, t_dep is ~1-2 Gyr
%due to the long time gas spends in the non-star foming phase, and the short time *only a
%small fraction* of the mass spends in a star froming cold phase. In contrast,
%the ISM in LARS8 is such that the disc is violently unstable, and much more mass is incoproprated
%into a number of dynamically dominating clouds

\begin{figure}
	\includegraphics[width=1\columnwidth]{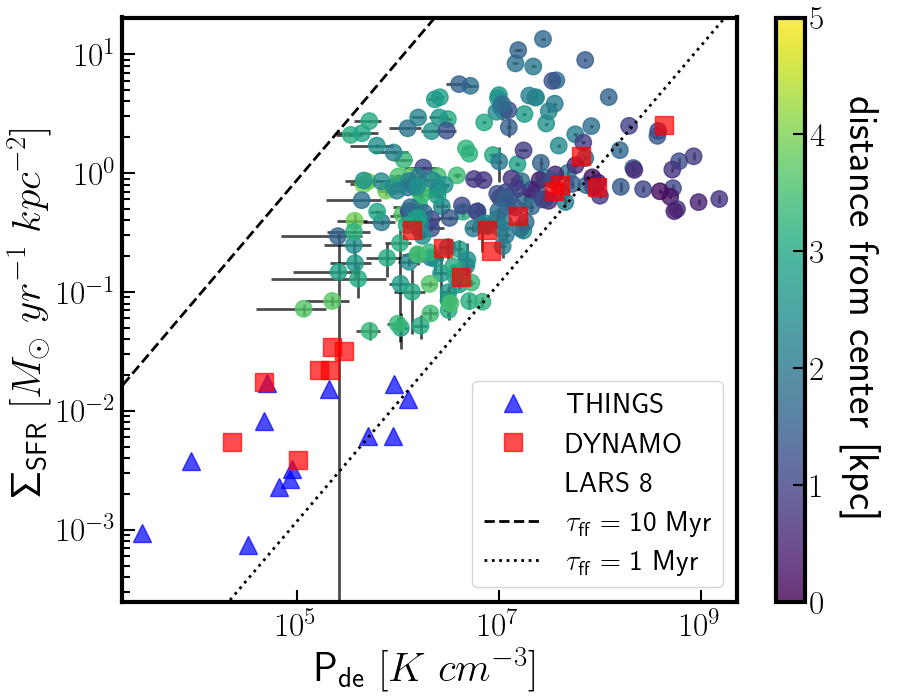}
    \caption{Star formation rate surface density (y-axis) versus dynamical equilibrium pressure (x-axis). Each point corresponds to a measurement in a $\sim$650\,pc sized region/pixel. The colors indicate the distance from the center.}
    \label{fig:pressure_sfr}
\end{figure}

% Pressure - SFR relation
We now test our observations
against models that are based on the assumption that star formation
is self-regulated through a balance between ISM pressure and
feedback. For example, \cite{Ostriker2011} and \cite{Kim2013}
predict in their semi-analytic models a (nearly) linear relationship
between the pressure and the star formation rate surface density:
$\Sigma_\mathrm{SFR}=4\ f^{-1}\ P_\mathrm{de}$. As described in
\cite{Fisher2019} the scaling factor $f$ can be determined from
$\sigma_\mathrm{gas}=0.366\ (\tau_\mathrm{ff}/\tau_\mathrm{depl})\ f$ \citep{Shetty2012}.
This leaves the free-fall timescale $\tau_\mathrm{ff}$ as the only
unknown. \cite{Krumholz2012} further estimate that
the range of $\tau_\mathrm{ff}$ should be between 1--10\,Myr for starburst galaxies,
i.e. in high density regimes.
In Figure \ref{fig:pressure_sfr} we plot the star formation rate densities against
pressure for LARS~8 and two comparison samples.
Each point in the plot represents an independent measurement (line-of-sight) that
we calculated from the mean value within 2x2 bins.
The dashed and dotted lines
indicate the model predictions for the above mentioned range in $\tau_{\mathsf{ff}}$
and a fixed gas depletion time of 300\,Myr that we typically find in the disc of LARS~8.
The Figure shows that the predicted linear relation does not
fit the data, we rather find evidence for a sub-linear trend,
similar to \cite{Fisher2019}. The slope in LARS~8, however,
seems even shallower, in particular in the low-pressure
regime. We conclude that in the outskirts of the observed disc 
the star formation is out of
equilibrium as
described in feedback-regulated star formation models, and is dictated by large scale instabilities instead.

% Cloud properties: alpha_vir versus Toomre Q
The importance of the large-scale environment for star formation
in LARS~8 is also reflected by the fact that the 
virial parameter of the identified diffuse molecular structures
(see Table \ref{tab:clumpprops}) has values that are roughly
identical to those found in Milky Way GMCs or normal disc galaxies
\citep{Sun2018}. Most clumps are found to be virialized
with $\alpha_\mathrm{vir}\sim1$ in which kinetic and gravitational
energy are roughly balanced.
This provides further evidence that on the scale of a few hundred parsec
the stars form in a roughly
uniform environment.
The high star formation rates observed in LARS~8 must thus be caused by
an increase in the number of clouds that are
triggered by large-scale gravitational instability (with low Toomre~$Q$).
Hence, the shorter gas depletion time scale -- or higher star formation efficiency --
observed in the outer disc does not imply that on our clump scales
the process of star formation is more efficient, but rather that the
formation of individual clumps is more efficient.

% Short note on difference NOEMA-single dish CO (2-1) from Puschnig+ (2020)
%
% Relate to LyAlpha escape (?)
%
% Impact of conversion factor on results
Next, we discuss how the choice of a fixed CO conversion factor $\alpha_\mathrm{CO}$
impacts our findings of the radial trend of \tdep\ and the Toomre~$Q$ instability of the disc.
We know from the MUSE data that there is a slight increase in metallicity
towards the center of the galaxy. Hence, application of a metallicity-dependent
conversion factor would only lead to a relatively lower
value of $\alpha_\mathrm{CO}$ in the center than in the outer part of the disc.
As a result, this would only exaggerate the observed trend
of decreasing molecular gas depletion with radius.
For the results of our disc stability analysis, the fixed conversion factor has only
minor impacts for two reasons.  First, the instabilities are mainly driven by the stellar component
(which is formally also shown in \citealt{Romeo2013}).
Second, it would only lead to slightly higher gas surface densities in the disc,
lowering support of the disc against collapse and thus resulting in even lower $Q$ values.

However, not only the metallicity impacts the conversion factor.
In infrared galaxies, but also in the centers of nearby galaxies, the nuclear zone is sometimes found to have
lower $\alpha_\mathrm{CO}$ caused by hotter molecular gas, thus higher velocity dispersion (which reduces the CO optical depth). This is
seen e.g. in NGC~6946 \citep{Meier2004}. If CO optical depths in the center of LARS~8
were systematically lower, we would need to use a lower conversion factor.
In the case of a typical ULIRG value ($\alpha_\mathrm{CO} \sim 1$), the central
depletion time would then drop from 1.35\,Gyr to 300\,Myr.
At the same time, this would
provide even further support against collapse in the central zone.
We plan to resolve this issue with observations of CO isotopologues
in a future study.

% Comparison Halpha - CO kinematics
%Levy et al. (2018) find in nearby spiral galaxies that ionized gas gives systematically lower rotation velocities than molecular gas.
%They argue that this is because in low-z spiral molecular gas is in a thin disc, where the ionized gas traces a thicker, more turbulent component. However, it is not clear if
%the same result holds for LARS 8,  that  has significantly higher surface densities of gas and star formation.  \cite{White2017} argue that for systems in dynamical
%equilibrium, which have large gas mass surface densities, the bulk of the gas will naturally have higher scale heights.

%%%%%%%%%%%%%%%%%%%%%%%%%%%%%%%%%%%%%%%%%%%%%%%%%%%%%%%%%%%%%%%%%%%%
%%%%%%%%%%%%%%%%%%%%%%%%%%%%%%%%%%%%%%%%%%%%%%%%%%%%%%%%%%%%%%%%%%%%
%%%%%%%%%%%%%%%%%%%%%%%%%%%%%%%%%%%%%%%%%%%%%%%%%%%%%%%%%%%%%%%%%%%%
%%%%%%%%%%%%%%%%%%%%%%%%%%%%%%%%%%%%%%%%%%%%%%%%%%%%%%%%%%%%%%%%%%%%
\section{Summary and Conclusion}\label{sec:summary}
We have obtained new high-resolution NOEMA CO~(2--1) and MUSE spectroscopy of
the $z\sim$0 massive, clumpy and gas-rich disc galaxy
LARS~8, drawn from the \textit{Lyman Alpha Reference Sample}.
The NOEMA data was used to study the diffuse molecular gas content and its kinematics
at a resolution of $\sim$400\,pc, while the MUSE data was used to derive extinction-corrected
star formation rates from H$\upalpha$ at a resolution of $\sim$600\,pc.
This enabled us -- together with readily available HST photometry -- to perform
a disc stability analysis using the Toomre~$Q$ criterion. The main
result is presented in Figure \ref{fig:toomreQ},
showing that the disc is highly unstable ($Q<<$1) over large scales.
On the other hand, the
center of LARS~8 was found to be stable ($Q>$1).

The NOEMA molecular data cube was further examined with \texttt{CPROPSTOO},
allowing us to identify and compute physical properties of 12 individual
molecular clumps (Table \ref{tab:clumpprops}). The clumps are
found to be virialized ($\alpha_\mathrm{vir}\sim$1) and they
follow the mass-size relation (Figure \ref{fig:mass_size_rel}).

We have further derived several physical parameters such as the
molecular gas depletion time, the molecular gas fraction and
the dynamical equilibrium pressure. Using our results from the
CO-based rotation curve (Figure \ref{fig:rotcurve}), all (surface) quantities could be corrected
for inclination effects. The radial (elliptical) profiles are
shown in Figure \ref{fig:radprof}. Of particular interest is
the smooth radial decline of the molecular gas depletion
time, ranging from more than 1\,Gyr in the center to $\sim$100\,Myr
in the outer disc. This trend is outstanding, as in normal
star forming galaxies the gas depletion time is observed to be
constant or even slightly increasing with radius.
These results lead to the following conclusions:

\begin{itemize}
    \item The disc of LARS~8 is highly unstable with $Q<<$1 and has relatively short gas depletion times.
    The identified diffuse molecular structures, however, are virialized and thus similar to GMCs in the Milky Way
    or nearby galaxies. Hence, the short gas depletion times in the disc cannot be explained by local (sub-kpc)
    effects such as a higher local star formation efficiency, but must be triggered by large-scale
    processes that cause the formation of more massive
    and \textit{denser}
    molecular clumps.
    The observed short gas depletion times observed in the disc must thus result from more
    dense gas being present on sub-clump scales, i.e. density PDFs shifted towards higher values.
    We argue that the high star formation rates observed in LARS~8 are the result
    of large-scale Toomre instabilities in the galaxy disc.

    \item The central region of LARS~8 is Toomre-stable, has the longest gas depletion time,
    lower gas fraction and a reduced star formation rate surface density. Given the fact
    that the stellar surface density (and thus the ISM pressure) is found to be highest
    in the center, we argue that some process must lower star formation in the
    central $\sim$500\,pc.
    From our dynamical analysis we find evidence that shear (and not feedback from SNe) is the main driving mechanism
    that suppresses the star formation in the center of LARS~8,
    as $\kappa$ increases by a factor of 4 from r$\sim$1\,kpc to r$\sim$0\,kpc.
    
\end{itemize}

%%%%%%%%%%%%%%%%%%%%%%%%%%%%%%%%%%%%%%%%%%%%%%%%%%%%%%%%%%%%%%%%%%%%
%%%%%%%%%%%%%%%%%%%%%%%%%%%%%%%%%%%%%%%%%%%%%%%%%%%%%%%%%%%%%%%%%%%%
%%%%%%%%%%%%%%%%%%%%%%%%%%%%%%%%%%%%%%%%%%%%%%%%%%%%%%%%%%%%%%%%%%%%
%%%%%%%%%%%%%%%%%%%%%%%%%%%%%%%%%%%%%%%%%%%%%%%%%%%%%%%%%%%%%%%%%%%%
\section*{Acknowledgements}
J.P. acknowledges funding from the European Research Council (ERC) under the European Union's
Horizon 2020 research and innovation programme (grant agreement No.726384/Empire).
M.H. is Fellow of the Knut and Alice Wallenberg Foundation.
O.A. acknowledges support from the Knut and Alice Wallenberg Foundation and from the Swedish Research Council (grant 2019-04659).
This research made use of Astropy, a community-developed core Python package for Astronomy \citep{Astropy}.

%%%%%%%%%%%%%%%%%%%%%%%%%%%%%%%%%%%%%%%%%%%%%%%%%%

\section*{Data Availability}
The MUSE raw data underlying this article are available in the ESO
public archive at http://archive.eso.org/
and can be accessed with the program ID 0101.B-0703(A).
The reduced MUSE data cube and the NOEMA data underlying
this article will be shared on reasonable request to the corresponding author.



\bibliographystyle{mnras}
\bibliography{clumpform} % if your bibtex file is called example.bib

%%%%%%%%%%%%%%%%%%%%%%%%%%%%%%%%%%%%%%%%%%%%%%%%%%

%%%%%%%%%%%%%%%%% APPENDICES %%%%%%%%%%%%%%%%%%%%%
\appendix

\section{Observed vs. modeled moment maps}

\begin{figure*}
	\includegraphics[width=\textwidth]{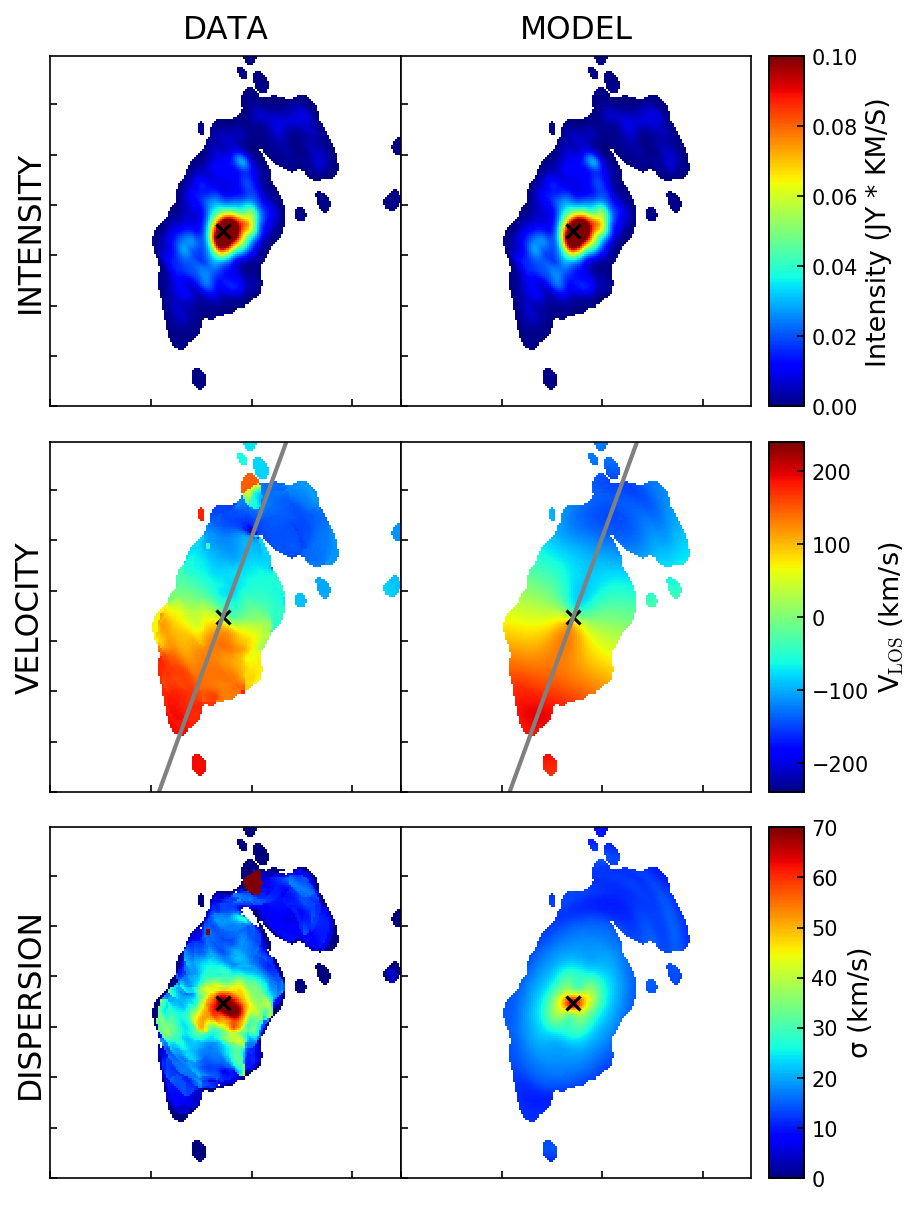}
    \caption{Observed (\textit{left column}) versus modeled (\textit{right column}) properties of LARS 8. The latter were produced with \texttt{$^{3D}$BAROLO}.}
    \label{fig:3dbarolo_obs_model}
\end{figure*}

%%%%%%%%%%%%%%%%%%%%%%%%%%%%%%%%%%%%%%%%%%%%%%%%%%

% Don't change these lines
\bsp	% typesetting comment
\label{lastpage}
\end{document}